\documentclass[12pt]{iopart}

\usepackage{iopams}

\usepackage{graphicx}
\usepackage{sidecap}
\usepackage{fancyhdr}
\begin{document}

\title{Engineered Quantum Dot Single Photon Sources}

\author{Sonia Buckley, Kelley Rivoire, Jelena Vu\v{c}kovi\'{c}}

\address{Center for Nanoscale Science and Technology, Stanford CA 94305}
\ead{bucklesm@stanford.edu}
\begin{abstract}
Fast, high efficiency, and low error single photon sources are required for  implementation of a number of quantum information processing applications. The fastest triggered single photon sources to date have been demonstrated using epitaxially grown semiconductor quantum dots (QDs), which can be conveniently integrated with optical microcavities.  Recent advances in QD technology, including demonstrations of high temperature and telecommunications wavelength single photon emission, have made QD single photon sources more practical. Here we discuss the applications of single photon sources and their various requirements, before reviewing the progress made on a quantum dot platform in meeting these requirements.
\end{abstract}

\tableofcontents

\section{Introduction to single photon sources}
\subsection{Introduction}
\begin{figure}[h!]
\includegraphics[width = 15cm]{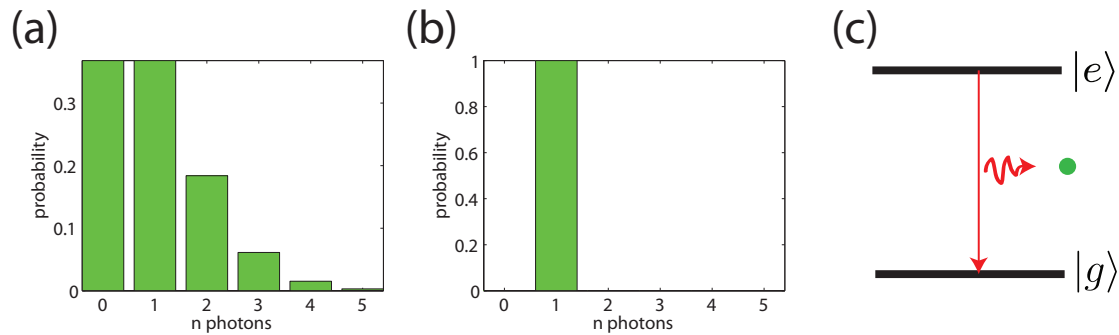}
\caption{Coherent (a) and Fock number (b) states with mean photon number of 1. n is the number of photons. (c) A two level system emitting a single photon.}
\label{fig:introduction}
\end{figure}

An ideal single photon source emits a single photon with a probability of 1 in response to an external trigger, and hence, has a probability of 0 to emit more or fewer than 1 photon. However, the probability of emitting a single photon cannot be 1 either for a coherent source of light (such as a laser), or for a thermal source, because both of these emit a distribution around a mean number of photons.  A coherent state of light has a Poisson distribution of photons with a mean photon number $|\alpha|^2$, written in the Fock state basis with $n$ as photon number as
  \begin{equation}
  |\alpha \rangle = e^{-\alpha^2/2}\displaystyle\sum_n\frac{\alpha^n}{n!}|n\rangle.
  \end{equation}
  This distribution is illustrated in Fig. \ref{fig:introduction} (a).  No matter how such a source is attenuated, there will always be some probability of obtaining photon numbers not equal to 1.  A single photon source, therefore, must emit light in a non-classical number state, called a Fock state. This is illustrated in Fig. \ref{fig:introduction} (b).  The two main types of single photon sources studied today use an atom or atom-like system, or a nonlinear material process such as spontaneous parametric downconversion (SPDC). Atom-like systems can be triggered to emit single photons on demand, while SPDC is by nature a random process, and can at best use another photon to `herald' the generation of a single photon.   For the remainder of this review we will discuss atom-like systems as single photon sources.

 An atom-like system is induced to emit a single photon either via optical or electrical excitation.  In the case of optical excitation, we start out with an incoming laser pulse in a coherent state, where the photon number follows a Poisson distribution. The atom is used to convert this into a single photon stream.  The atom can be modeled as a two level system with a ground state $|g \rangle$ and an excited state $|e \rangle$, illustrated in Fig. \ref{fig:introduction} (c). The atom in the excited state $|e\rangle$ emits a single photon via spontaneous emission from $|e\rangle$ to $|g \rangle$. Once it decays from the excited state to the ground state it can no longer re-emit a photon until it is excited again.  This tendency to emit single photons separated in time is called anti-bunching. \\

The first optical wavelength single photon sources were demonstrated in the late 1970s using a beam of sodium atoms excited by a continuous wave laser \cite{kimble_photon_1977}.  Solid-state systems were first investigated as single photon sources in the 1990's, with the first demonstration of anti-bunching performed using a single dye molecule \cite{basche_photon_1992}. This was followed by other solid state systems such as nitrogen vacancy centers \cite{kurtsiefer_stable_2000} in diamond and CdSe quantum dots \cite{michler_quantum_2000-2}.  The first epitaxial self-assembled semiconductor quantum dot (QD) used as a single photon sources appeared around the same time \cite{michler_quantum_2000-1}, and since then there has been an explosion of work on the topic.  QDs are an excellent source of single photons, with perhaps their biggest advantage being the ease of integration with optical microcavities, which can be fabricated around them.  In this section we will begin with a discussion of applications for a single photon source, before describing the properties of an ideal single photon source for these applications.  We will also briefly discuss different solid state emitter systems before focusing on semiconductor quantum dots.  For more details, good references on single photon sources from a variety of perspectives can be found in \cite{scully_quantum_1997}, \cite{lounis_single-photon_2005}, \cite{michler_single_2010} and \cite{santori_single-photon_2010}. For a briefer overview of semiconductor quantum light emitters, see a review by Andrew Shields \cite{shields_semiconductor_2007}.

\subsection{Applications}

The macroscopic objects we experience in our daily lives appear to follow a set of deterministic rules.  At the single photon level however, these rules no longer apply, and startling quantum mechanical effects can be observed. Various useful applications of these non-intuitive effects are being studied.  Quantum key distribution and quantum information processing protocols such as linear optical quantum computing take advantage of the fact that quantum mechanical objects can exist in a superposition of states that collapses when observed \cite{nielsen_quantum_2000}.  In the case of quantum key distribution, this makes it impossible to `eavesdrop' on a secure connection without being observed, since the state will collapse upon observation by the eavesdropper.  In the case of quantum computing, this can be exploited to solve certain problems significantly faster than with a classical computer \cite{nielsen_quantum_2000}.  Other applications for single photons take advantage of different properties, such as the elimination of shot noise in low signal measurements due to the squeezed nature of single photons \cite{walls_squeezed_1983, xiao_precision_1987}.  Single photons can also be used to create multi-entangled states, e.g Greenberger-Horne-Zeilinger states \cite{greenberger_multiparticle_1993}, which can be used for greater measurement accuracy e.g in beating the diffraction limit for a particular radiation wavelength of light.  Below, we go into more detail on a few different applications. All of these applications benefit from increased speed, which leads to increased data rates.

\subsubsection{Quantum key distribution}
Quantum key distribution (QKD) is a method to secretly exchange a key between two distant partners, traditionally referred to as Alice and Bob, in the presence of an adversary, referred to as Eve. The advantage of this over classical key distribution methods stems from the quantum mechanical observer effect, which refers to the fact that it is impossible to directly measure a quantum mechanical state without changing it.  This means that given a perfect experimental system, Eve cannot intercept Alice and Bob's quantum key without being noticed. Various different schemes for implementation have been proposed \cite{gisin_quantum_2002}.  One of the first protocols was proposed by Bennett and Brassard in 1984 \cite{bennett_quantum_1984} and is called the BB84 protocol.  This uses four different polarization states in two conjugate bases:  a straight basis with horizontal $|H\rangle$ and vertical $|V\rangle$ basis states, and a diagonal (45$^\circ$) basis with $|F\rangle$ and $|S\rangle$ as basis states. Alice sends individual photons with random polarization states to Bob, who measures them using one of the two bases, also at random.  They can now publicly exchange information on which bases they have used since Eve does not know what result they got. Whether or not their results are correlated will depend on whether they chose the same or different bases. The bits measured in different bases can be thrown away, and the remaining bits used to construct a key.  In principle, Alice and Bob both know this key exactly, although it was created randomly. In practice, experimental errors or eavesdropping may mean that this key has errors.  Classical error correcting algorithms can then be used to correct these errors \cite{bennett_privacy_1988}.

The eavesdropper Eve has to intercept and measure the photons, and to avoid arousing Bob's suspicion she must send another photon in its place.  According to the no-cloning theorem, no matter what technology Eve has she cannot produce a perfect copy of an unknown quantum system \cite{wootters_single_1982}.   In order to avoid being noticed, she could send a different photon (which will not be a copy of the one she received) to Bob; however, in this case she will increase Bob's error rate and risks being noticed.  If a weak laser signal is used in quantum cryptography, it will  sometimes send multiple photons, in which case Eve can simply intercept one of these multiple photons and extract information.  This is therefore less secure than a true single photon source \cite{lutkenhaus_security_2000}.

Using an attenuated laser signal with average photon number much less than one (and therefore low probability for $>$1 photon) is significantly cheaper and more convenient experimentally than using a true single photon source. Protocols to increase the security of QKD
based on these sources have been devised, for example altered protocols that defeat the number splitting attack have been designed \cite{sangouard_what_2012} and commercially implemented.  These protocols lessen the need for a true single photon source.  The need for a real single photon source for QKD can therefore be called into question \cite{sangouard_what_2012}.  Single photon
sources do however have some advantages over attenuated lasers. The attenuated laser protocols are source dependent, leaving the source open to attack or misuse by an unknowledgeable operator.  Additionally, in order to send faint signals over long distances, a quantum repeater is necessary. This requires a true single photon source for operation \cite{duan_long-distance_2001}.  The first demonstration of QKD
with a pulsed true single photon source was demonstrated with single photons from nitrogen-vacancy centers in 2002 \cite{beveratos_single_2002}.
\subsubsection{Linear optical quantum computing}
Photons are very good as quantum bits due to their ability to travel long distances, negligible decoherence and the fact that encoding can be implemented in any of several degrees of freedom (for example polarization, time bin or path \cite{obrien_optical_2007}).  However, they interact only very weakly, which makes realizing the logic gates needed for a quantum computation scheme challenging.  The controlled-NOT or CNOT gate has been shown to be a universal logic gate for quantum computers \cite{nielsen_quantum_2000}. By composing CNOT gates, other unitary transformations can be built.  The CNOT gate transformation acts as
\begin{equation}
\mathrm{CNOT}:|a,b\rangle\rightarrow|a,a\oplus b\rangle
\end{equation}
where $a \oplus b$ denotes addition modulo 2. This logic operation is inherently nonlinear because the state of one quantum particle must be able to control the state of the other.  Knill et al. \cite{knill_scheme_2001} have showed that a CNOT gate can be implemented using only linear optics and photon counting detection.  This means that in principle, a quantum computer could be realized using photons as qubits despite very weak photon-photon interaction. However, there are strict requirements on the single photon sources needed for this protocol: very low error rates and high efficiencies are needed. Reports of fault tolerances vary \cite{obrien_optical_2007}. It has been shown that if all other components are perfect, quantum computation is possible if the product of source and detector efficiency is $>$ 2/3 \cite{varnava_how_2008}. A more recent paper showed that source efficiencies of 0.9 with g$^{(2)}(0) < 0.07$ \cite{jennewein_single-photon_2011}.  An additional stringent requirement is that the photons must undergo quantum interference on beamsplitters, which means that the photons must be indistinguishable (discussed in section \ref{section:indistinguishable}).  The data or bit rate will also be limited by the single photon source speed, and so a fast source is a requirement.

\subsubsection{Quantum metrology}
The Heisenberg uncertainty relation puts a fundamental limit on the precision of a measurement.  Most standard measurement techniques, however, do not reach this limit and are instead limited by otherwise avoidable sources of error stemming from non-optimal measurement strategies \cite{giovannetti_quantum-enhanced_2004}.  For example, a coherent state distributes its quantum-mechanical uncertainties equally between position and momentum, and the relative uncertainty in phase and amplitude are roughly equal.  By using a squeezed state, the uncertainty (noise) in phase, amplitude or a general quadrature can be reduced (while the uncertainty will be increased elsewhere).  By choosing a state with low noise in the desired quadrature, an optimal measurement strategy can be devised. Fock states, such as the $n$=1 single photon state, are squeezed states of light, with a fixed number of photons but indeterminate phase.

Shot noise is a good example of noise arising from a non-optimal measurement strategy; this noise is $\sqrt{N}$ for coherent light with a mean number of $N$ photons, while for a Fock state, such as a single photon state, shot noise is completely eliminated. This elimination of shot noise will allow better measurements of weak absorptions; when a coherent source is used shot noise puts a limit on the weakest absorption that can be measured.  A perfect single-photon source associated with perfect detection would give access to arbitrarily small absorptions, impossible to measure with a laser source because of photon noise \cite{lounis_single-photon_2005}.

Another place where quantum effects can be employed to reduce measurement uncertainty is in increasing minimum feature size that can be resolved using a particular wavelength source.  This minimum feature size is defined by the Rayleigh diffraction limit.  Reducing the wavelength will reduce this minimum feature size; however, in practice, shorter wavelengths are sometimes difficult to generate and focus, or lead to unwanted damage to the sample being measured. To illustrate how this can be overcome, let's consider a simple quantum-mechanical object described with a plane wave-like wavefunction; the quantum mechanical wavelength is $\lambda_{qm} = 2\pi\hbar/p$, where $p$ is the object's momentum.  For a single photon, the radiation wavelength $\lambda_{rad} = 2\pi c/\omega$, and the momentum is $p = E/c$, where $E = \hbar\omega$ is the energy of the photon.  Therefore its quantum mechanical wavelength $\lambda_{qm}$ is equal to its radiation wavelength $\lambda_{rad}$.  For a two photon state, the momentum would be two times larger (since E = $2\hbar\omega$), and thus $\lambda_{qm}$ would be two times smaller than for a single photon with the same radiation wavelength: $\lambda_{qm} = \lambda_{rad}/2$.  Using higher photon number states would lead to further decreases in the wavelength $\lambda_{qm}$. The most important application for this would be lithography to reduce the minimum feature size \cite{boto_quantum_2000}, since the diffraction limit is determined by $\lambda_{qm}$.  Single photons can be used to create these multi-photon states by interference at a beamsplitter \cite{fattal_entanglement_2004}.

Similarly, precise measurement of an object usually relies on a measurement of the time it takes for light signals to travel from that object to some known reference points. For single photons the time of arrival of each of the photons will have a spread 1/$\Delta \omega$, where $\Delta \omega$ is the bandwidth. If one measures an average arrival time for N single photon pulses, the error in the travel time will be $\frac{1}{\Delta\omega\sqrt{N}}$. However, if one generates an entangled state with $N$ photons and measures its arrival time, the error would be $N$ times smaller than for a single photon.  This is because such an N photon state effectively has N times higher frequency \cite{giovannetti_quantum-enhanced_2001}.  Therefore the entanglement gives an overall gain of $\sqrt{N}$ relative to the employment of N individual photons (i.e. a classical approach of averaging N arrival times) \cite{giovannetti_quantum-enhanced_2001}.

\subsubsection{Single photon quantum memory}
Photons are ideal for carrying quantum information: they can travel long distances with low transmission losses and experience minimal decoherence.  However, they are difficult to store for a long time. In order to implement a quantum memory for quantum information transmitted via photons, it is necessary to map the quantum state of the light pulse to another medium.  The spin of an electron (or hole) is an excellent candidate for a stationary or storage qubit. Such a quantum memory is essential for the development of many devices in quantum information processing, including a synchronization tool that matches various processes within a quantum computer, and for the implementation of quantum repeaters, which in turn are necessary for long distance quantum communication \cite{lvovsky_optical_2009, simon_quantum_2010}.
Proposals for quantum memories include ensembles of atoms \cite{kozhekin_quantum_2000}, solid state atomic ensembles such as rare earth dopants in glass \cite{tittel_photon-echo_2010}, single atoms  \cite{cirac_quantum_1997,maitre_quantum_1997} and single impurities in solids, such as NV centers in diamond \cite{togan_quantum_2010} and charged epitaxially grown QDs \cite{kroutvar_optically_2004}.

\subsection{Brief description of requirements}

The aforementioned applications all imply various and differing requirements on generated single photons.  Here we will describe some of these requirements.

\subsubsection{Operating temperature}
Solid state atom-like emitters in general exhibit phonon induced linewidth broadening, and at high temperatures excited state transitions will often overlap, leading to loss of single photon character.  The narrowest linewiths are observed at cryogenic temperatures, and solid state single photon sources will experience the least dephasing and demonstrate the highest indistinguishabilities at these low temperatures.  In addition, for many epitaxial QD systems, the thermal energy exceeds the confinement potential at higher temperatures, and the QDs will stop luminescing as the temperature is raised \cite{le_ru_temperature_2003}.  For practical applications, it is desirable to have a single photon source that works at room temperature.  Liquid nitrogen cooling (available above 77 K) is also significantly more practical than liquid helium cooling.  While many single photon sources have been demonstrated at room temperature, all of these have shown significant linewidth broadening \cite{beveratos_room_2002, lounis_single_2000, michler_quantum_2000-2, fedorych_room_2012}.

\subsubsection{Wavelength}
Ideally, a single photon source would be a narrow linewidth emitter tunable over a very broad frequency range, or else a highly efficient method for frequency conversion to arbitrary wavelengths would be necessary.  This would allow selection of the optimal wavelength for a particular application.  Additionally, with precise wavlength control, correcting for the discrepancy in the emitter transition energies resulting from inhomogeneous broadening would also be possible, allowing interaction between different nodes in a quantum network, and allowing interference between single photons from different emitters, e.g. for formation of multi-photon entangled states. However, this broad tunability has yet to be realized in a practical source.  For quantum cryptography, for example, it is desirable to transmit single photons over long distances with minimal losses.  Silica telecommunications wavelength fibers have two main transmission windows at 1320 (O-band) and 1550 (C-band) nm. However, photon detectors in these wavelength ranges are typically made of InGaAs and currently have significantly worse performance than the Si photodetectors, which have peak detection efficiency in the visible range at around 750 nm, with detection extending out to around 1000 nm. For applications where high detection efficiency is important, emitters in this wavelength range are more desirable.  Frequency conversion and advances in detectors in the telecom wavelengths will be discussed in section \ref{section:currentfutureresearch}. Emitters in the blue and UV part of the spectrum are also potentially interesting for QKD, as the emitters and receivers could be smaller for this wavelength range, and plastic fibers have transmission minima there.

\subsubsection{Speed}

 The speed of a single photon source is determined by its emission lifetime $\tau_f = 1/\Gamma_0$, a characteristic of the emitter, and $\Gamma_0$ is the spontaneous emission rate of the emitter. For a quantum emitter in a uniform medium with refractive index n, the spontaneous emission rate is completely determined by the transition frequency, $\omega$, and by the transition dipole moment, $\mu_{eg}$, between ground and excited states
\begin{equation}
\Gamma_0 = \frac{4}{3 n}\frac{\mu_{eg}^2}{4 \pi \epsilon_0\hbar}\left( \frac{\omega}{c} \right)^3.
\end{equation}
This spontaneous emission lifetime will be modified depending on the local photonic density of states in the vicinity of the emitter.  We will discuss this further in section \ref{section:microcav}.  In the ideal system the linewidth of the emitter will be Fourier-transform (lifetime) limited. In practice, however, solid state systems are often excited via incoherent pumping (see section \ref{section:above-band}) and thus the speed of relaxation from higher energy levels to the excited level state must be taken into account.  This can lead to both longer effective lifetimes and jitter in the emission time of a single photon pulse.

High speeds are desirable in order to achieve high data rates desired for quantum information processing; speeds of at least 1-10 Gbps are desirable for applications such as QKD.  In addition, the time taken to perform tasks such as creation of N-photon entangled states from single photons increases scales as $t^N$. A recent experiment created an 8 photon entangled state from 4 entangled photon pairs \cite{yao_observation_2012} at a rate of 9 detected 8-photon states per hour. Increasing the rate of generation of single photons and entangled photon pairs would help greatly to increase the speed of higher entangled photon states.

\subsubsection{Efficiency}
The efficiency of a single photon source is the fraction of triggers leading to the generation of a single photon.  Very low error rates are necessary for QIP, and efficiencies of greater than 99 \% are desired for all-optical quantum computing \cite{kiraz_quantum-dot_2004}, although it has been shown that if all other components are perfect, quantum computation is possible if the product of source and detector efficiency is $>$ 2/3 \cite{varnava_how_2008}. For QKD, the security of the connection will improve the higher this efficiency and the lower the error rate; therefore very high efficiencies are also necessary for this application.

\subsubsection{g$^{(2)}(\tau)$}

The most important measurement for verifying that a source is indeed emitting single photons is the g$^2(\tau)$ or photon intensity autocorrelation measurement.  Verifying that a source exhibits `anti-bunching' with g$^{(2)}(0)=0$ qualifies it as a bona-fide single photon source.  Demonstrating g$^{(2)}(0) < 1$ is an entirely non-classical result and proves the quantum nature of the radiation.  Here, we will define  g$^{(2)}(\tau)$, while in section \ref{section:measurementofg20} we will  discuss how it can be practically measured. More detail can be found in quantum optics books, e.g. \cite{scully_quantum_1997}.
The first-order coherence function is defined as
\begin{equation}
g^{(1)}(\tau) = \frac{\langle \hat{a}^{\dagger}(t) \hat{a}(t+\tau)\rangle}{\langle \hat{a}^{\dagger}(t) \hat{a}(t) \rangle}
\end{equation}
and second order coherence as:
\begin{equation}
g^{(2)}(\tau) = \frac{\langle \hat{a}^{\dagger}(t) \hat{a}^{\dagger}(t+\tau) \hat{a}(t+\tau) \hat{a}(t)\rangle}{\langle \hat{a}^{\dagger}(t) \hat{a}(t) \rangle^2}
\end{equation}
We can see that the first-order correlation is insensitive to the photon statistics, since this expression only depends on the average photon number $\langle n \rangle = \langle \hat{a}^{\dagger}\hat{a}\rangle$.  In other words, spectrally-filtered thermal light and coherent light with the same average photon number exhibit the same degree of first-order coherence. This first order coherence determines the coherence length of the source. In contrast, the second-order coherence distinguishes between the different type of light fields. For a number state $|n\rangle$ of light,
\begin{equation}
g^{(2)}(0) =  \frac{\langle n|\hat{a}^{\dagger} \hat{a}^{\dagger} \hat{a} \hat{a}|n \rangle}{\langle n| \hat{a}^{\dagger}(t) \hat{a}(t)|n \rangle^2} = 1 - \frac{1}{n}
\end{equation}
For a true single-photon source $(n = 1)$, $g^{(2)}(0) = 0$. We state the results for coherent light (e.g. laser light)
\begin{equation}
g^{(2)}(0) =  \frac{\langle \alpha|\hat{a}^{\dagger} \hat{a}^{\dagger} \hat{a} \hat{a}|\alpha \rangle}{\langle \alpha| \hat{a}^{\dagger}(t) \hat{a}(t)|\alpha \rangle^2} = 1
\end{equation}
and for thermal light
\begin{equation}
g^{(2)}(0) =  1+\frac{\left( \Delta n \right)^2 - \langle n \rangle}{\langle n \rangle^2} = 2
\end{equation}
It is clear from the above equations that a single photon state can be distinguished from either coherent of thermal light by measuring $g^{(2)}(0) < 1$, and the presence of a single quantum emitter can be confirmed by measuring $g^{(2)}(0) < 1/2$.  In practice, measuring $g^{(2)}(0) < 1/2$  indicates the presence of the $n=1$ Fock state. For many applications, a low value of g$^{(2)}(0)$ is very important; e.g. for QKD multi-photon generation decreases the security of the encryption.

\subsubsection{Indistinguishability}
\label{section:indistinguishable}
Fearn and Loudon \cite{fearn_quantum_1987} and Hong et al \cite{hong_measurement_1987} pointed out that two photons incident at the same time on the two input ports of a 50\%/50\% beam splitter interfere in such a way that they both exit from one of the output ports. This effect is a consequence of the Bose-Einstein statistics followed by photons. The two photons `bunch', i.e. they always both exit through the same port of the beam splitter. Therefore, when the delay between the two incoming photons is varied, the rate of coincidences on the two output detectors drops for zero delay due to this bunching \cite{santori_indistinguishable_2002}. In order to give rise to a fully destructive interference, the two photons must be completely indistinguishable, i.e. they must be in exactly the same mode. Indistinguishability is important for linear quantum computing and other quantum information processing applications, which rely on interference between two single photons.  Additionally, quantum repeaters rely on the indistinguishability of photons, which means that for sending photons over long distances it may be necessary to have a high degree of indistinguishability.

Although pure spontaneous emission by an ideal, resonantly excited two-level system leads to perfectly indistinguishable photons, this indistinguishability can be lost by dephasing and spectral diffusion in the system, due to the fast and slow fluctuations of the transition frequency. If the spectrum of a single-photon source is Fourier-transform-limited, i.e. if each photon can be described by the same coherent wavepacket at the same frequency and polarization state, two photons will be indistinguishable. In many cases, however, the spectrum of a source is broader than the Fourier-transform of the time-profile of the emitted pulse. This broadening, arising from fluctuations of the optical resonance frequency, can be described as dephasing or spectral diffusion; such fluctuations impact the properties of photon wavepackets emitted at different instants of time, thus leading to distinguishability between successively emitted photons.  Dephasing causes the loss of coherence due to many �collision� events with a bath, leading to a gradual loss of phase with the dephasing (or decoherence) time, $T_2$, shorter than twice the fluorescence lifetime, $T_1$ \cite{santori_indistinguishable_2002}:
\begin{equation}
\frac{1}{T_2} = \frac{1}{2T_1}+\frac{1}{T_2^*},
\end{equation}
where $T_2^*$ characterizes pure dephasing processes arising from interactions with the bath.
In the absence of slow spectral diffusion, the resulting frequency linewidth (full-width at half-maximum),
\begin{equation}
\Delta\nu = \frac{1}{2 \pi T_1}+\frac{1}{\pi T_2^*},
\end{equation}
takes its minimum possible value only when dephasing is negligible, i.e. when $T_2^* = \inf$. In that case, one has a lifetime-limited linewidth.

For most systems in condensed matter at room temperature, the dephasing time is shorter by several orders of magnitude than the excited state lifetime, i.e. the linewidth is very far from being lifetime-limited.  In other words, indistinguishability requires lifetime-limited sources. In these cases, first-order coherence measurements can be applied to characterize their coherence length and the coherence time $T_1$.  Moreover, most single photon sources are not resonantly excited,  but they employ simpler above-resonant and quasi-resonant excitation methods, as described in section \ref{section:above-band}. In this case, a time jitter is introduced in the generation of single photons, resulting from the relaxation time of carriers from higher states. Such timing jitter additionally degrades the photon indistinguishability.

\subsubsection{Polarization}
A single-photon source that emits in a specific polarization is important for most applications.  The polarization is determined by the microscopic nature of the emitter (the orientation of its dipole moment) and by the way it is coupled to the emission mode.  A single self-assembled InAs/GaAs QD. for example, has two degenerate, orthogonally polarized one-exciton states from which the emission can be collected, and there will be thus no polarization preference. This changes when the emitter is coupled to a cavity - in this case due to the Purcell effect (see section \ref{section:microcav}) emission will occur preferentially into the cavity mode, which in general is strongly polarized.

    \subsection{Single emitter single photon sources, brief comparison}
Here we briefly describe and compare a few different single emitter single photon sources before focusing on quantum dot single photon sources.  For a more in depth review of other single photon sources see \cite{lounis_single-photon_2005}.

        \subsubsection{Atoms and ions}
        In comparison to solid state systems, atoms provide a very clean two level system.  They have purely electronic eigenstates with hyperfine structure.  In the atom and ion traps in which cavity QED and single photon experiments are done, the atoms have very narrow, lifetime limited linewidths \cite{kuhn_deterministic_2002, hijlkema_single-photon_2007}. Also unlike solid state emitters, the atomic states are perfectly reproducible and well-known, as all atoms are exactly the same. Excitation schemes for atoms and ions often rely on multi-step processes between known levels. The disadvantage of atoms is that the atomic systems are large and bulky and experiments tend to be complex. Typical radiative lifetimes of allowed atomic transitions are about 30 ns, corresponding to a linewidth of a few megahertz. This long lifetime limits the rate of generation of single photons.  For a reviews about using atoms for quantum information processing see ref. \cite{monroe_quantum_2002} and \cite{buluta_natural_2011} (comparing natural and artificial atoms).

\subsubsection{Molecules}
Molecules were the first solid state system observed to emit single photons \cite{basche_photon_1992} and also one of the first single photon sources to operate at room temperature \cite{lounis_single_2000}.  Due to their more complicated geometries and unlike atoms, molecules have vibrational states in addition to electronic states, which broaden the electronic states via the additional vibrations and phonons.  At very low temperatures, however, the lowest-frequency transition connecting the ground vibrational states of the ground and excited electronic states is a very narrow line, called the zero-phonon line (ZPL). The spectrum of the molecule at low temperature will be a narrow ZPL with other broader lines (shifted to the red with respect to the ZPL) corresponding to transitions between vibrational levels.  For indistinguishable photons, only photons from the ZPL can be accepted.  Molecules are strongly influenced by their environment, and due to environmental fluctuations, all molecules and molecular states will not be exactly alike. Molecular photostability is also a serious issue, due to the many photochemical processes that can occur, especially at room temperature and in an oxygen rich environment \cite{fleury_high_1998}. Blinking, a process in which fluorescent emission stops after applying the pump beam for a certain amount of time, and occurs due to the presence of a dark state, is also a serious issue with molecules, and can be either recoverable or non-recoverable \cite{zondervan_photoblinking_2003}.  Molecules can be positioned with respect to optical cavities, and enhancement of emission from a single molecule using nanoantennae has been demonstrated \cite{kinkhabwala_large_2009-1}, while coupling of molecules to photonic crystal cavities has also been shown \cite{rivoire_lithographic_2009}.

\subsubsection{Color centers}
Color centers are defects of insulating inorganic crystals, which localize electronic states generating a level structure that leads to fluorescence.  Although a variety of color centers have been studied, the most successful defect for quantum optics applications so far has been the nitrogen vacancy (NV) center in diamond \cite{jelezko_single_2006}.  This is also the first solid state emitter to be solid in turn-key commercial single photon source, recently available from Quantum Communications Victoria.  In addition to being the one of the first single photon sources to operate at room temperature \cite{gruber_scanning_1997}, it possesses interesting spin properties.  It consists of a carbon vacancy next to a nitrogen defect with a trapped electron (although a neutral version also exists, it does not possess the same spin coherence properties as the NV$^-$). The photoluminescence of the nitrogen vacancy center has a weak zero phonon line (ZPL) at 637 nm with a broad phonon side band (extending from 637 to 720 nm).  It is still visible at room temperature thanks to the stiffness of the diamond lattice, although the ZPL is weaker and broader at higher temperature.  The lifetime of the NV center is around 12 ns in bulk diamond.  Proximity to etched surfaces also damages the properties of the NV centers and changes this lifetime, which is problematic for coupling them to optical cavities or for using diamond nanocrystals.  There is also significant spectral diffusion.   It is also difficult to fabricate optical structures in the diamond substrate, although coupling of the ZPL of NV centers in diamond has been demonstrated \cite{faraon_resonant_2011}.  To overcome some of these shortcomings, a search for the optimal defect center is ongoing, and candidates such as defects in SiC \cite{koehl_room_2011} and other tetrahedrally coordinated semiconductors are being studied \cite{weber_quantum_2010, wrachtrup_defect_2010}.

        \subsubsection{Colloidal quantum dots}
        Colloidal quantum dots and semiconductor nanocrystals have size-dependent wavelength tunable emission most commonly situated in the visible part of the spectrum, and, due to their broad absorption continuum above the exciton transition, they can be excited with a variety of sources.   Anti-bunching from this system was first observed in 2000 \cite{lounis_photon_2000}, even at room temperature \cite{michler_quantum_2000-2}.  Their absorption and emission properties are similar to molecules. Nanocrystals are much more photostable than organic molecules under similar conditions. They are easy to manipulate and to couple to efficient collecting optics in a room-temperature microscope and have better stability than single organic chromophores. Their small size leads to localization of discrete electronic states. The spectrum is a single line (ZPL), with a weak phonon sideband. This ZPL is strongly broadened by dephasing and spectral diffusion and is thus very far from lifetime-limited. At low temperatures, this narrows down significantly, but never reaches the lifetime limit, probably because of spectral diffusion. This spectral diffusion and the very long luminescence lifetime, $\sim$ 20 ns \cite{labeau_temperature_2003}, are two weak points of nanocrystals for low-temperature applications as single-photon sources. Like with molecules, a serious limit to their practical application is blinking \cite{kuno_/off_2001}, although work on suppressing this blinking is ongoing \cite{mahler_towards_2008}.

        \subsubsection{Epitaxial semiconductor quantum dots}
Epitaxially grown semiconductor QDs have excellent optical stability compared to other solid state systems.  They are extremely bright, and have the advantage that they are easily integrated with other semiconductor structures and fabrication techniques, e.g. electrical control and optical microcavities.  Linewidths can be lifetime limited at cryogenic temperatures and are on the order of GHz.  We will discuss these and their other properties in the next section.  Good references on the subject of these QDs can be found in \cite{michler_single_2010, henneberger_semiconductor_2008}.

\subsection{Summary of single photon souces}

A summary of the properties of atom-like single photon sources is shown in Table 1.  In some cases properties for the most common emitters have been inserted.  All of the properties listed may not be available for the same system, e.g. lifetime and linewidth may change significantly from cryogenic temperatures to room temperature, even for systems that still exhibit anti-bunching at room temperature.

\begin{table}[h!]
\label{table:SPS}
  \begin{tabular}[h!]{ | p{2cm} || l | p{1.2cm} | p{1.5cm} |p{5cm} |p{1.5cm}| }
  \hline
    Emitter & $\lambda$ (nm) & $\tau$ (ns) & $T_{max}$ (K) & Comments & ref \\
    \hline \hline
    Atoms & * & $\sim$30 & ** & long coherence time & \cite{buluta_natural_2011} \\ \hline
    Ions & * & $\sim$30 & ** & long coherence time & \cite{buluta_natural_2011}\\ \hline
    Molecules & visible & $\sim$1-5 & Room T &  & \cite{lounis_single_2000} \\ \hline
    NV center & 640-720 & $\sim$12 & Room T & other defect centers in diamond being investigated & \cite{jelezko_single_2006} \\ \hline
    Colloidal QDs & 460-660 & $\sim$20-30 & Room T & for CdSe/ZnS system & \cite{lounis_photon_2000, michler_quantum_2000} \\ \hline
    Epitaxial QDs & 250-1550 & $\sim$0.1-10 & 40 - Room T & lifetime, wavelength and max T varies significantly with material (see table 2) & \cite{michler_single_2010}\\
     \hline
    \end{tabular}
    \label{Table:extinction}
\caption{Comparison of solid state single photon sources and typical properties. *discrete transition wavelengths depend on the emitter. **operated in room T vacuum system with laser cooling.}
\end{table}

\section{Introduction to quantum dots as single photon sources}

In this section, we will focus specifically on epitaxially grown QDs and their properties.

\subsection{Band structure}

\begin{figure}[h!]
\includegraphics[width = 15cm]{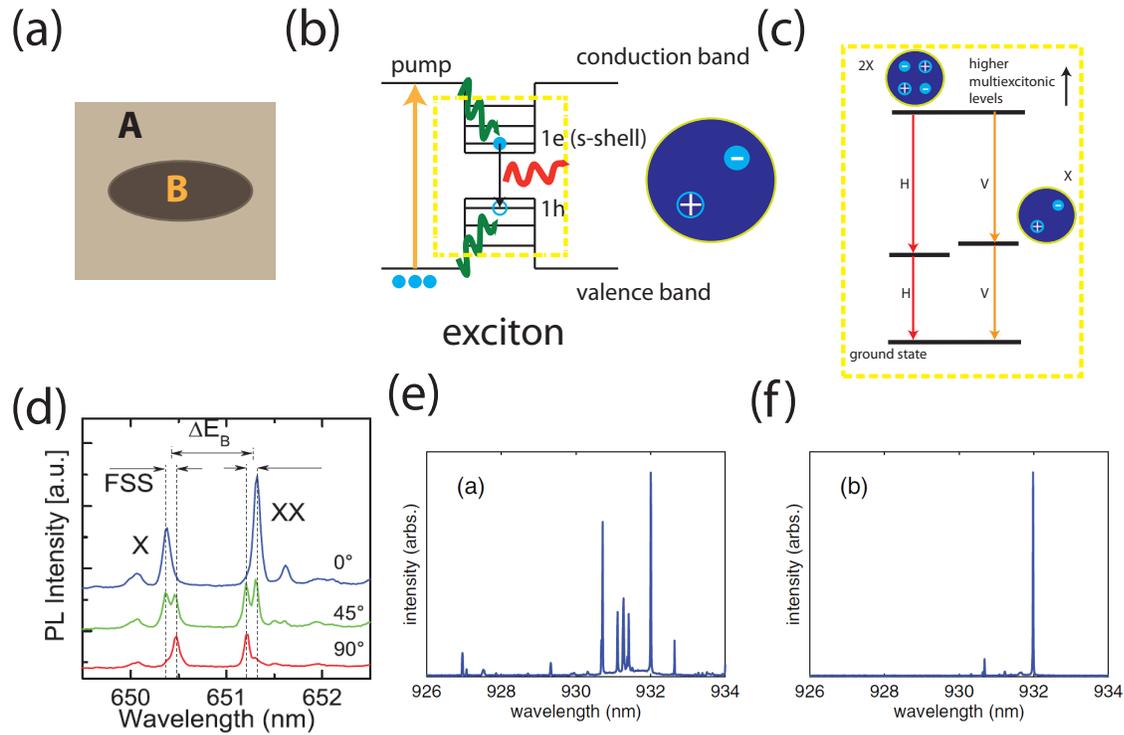}
\caption{(a) A QD consists of a small, nanoscale island of lower band gap semiconductor (B) embedded in a higher band gap semiconductor (A). (b) 1D diagram of the electronic structure of the QD. Incoherent pumping is shown and emission from the exciton state. (c) The level structure and fine structure splitting present for biexciton and exciton. (d) Fine structure splitting present in InP/InGaP QD. Spectrum shown for polarizations at 0$^\circ$, 45$^\circ$ and 90$^\circ$. Reproduced with permission from ref. \cite{ugur_single-photon_2012}. The spectrum of InAs/GaAs QD under (e) above band and (f) resonant excitation. In (f), the excitation laser is tuned to the higher order transition inside a QD, while in (e), the excitation laser frequency is above the GaAs band gap. (e)-(f) reproduced with permission from ref. \cite{santori_single-photon_2004}.}
\label{fig:electronic_structure}
\end{figure}

A QD consists of a lower band gap semiconductor (B) embedded in a higher band gap semiconductor (A) (Fig. \ref{fig:electronic_structure} (a)).  This leads to a three-dimensional electronic confinement due to the band offsets, and is illustrated in 1D in Fig. \ref{fig:electronic_structure} (b).  Initially, electrons are present in the valence band and holes in the conduction band. Optical or electrical excitation can cause an electron to be excited to the conduction band, leaving a hole in the valence band.  These electron-hole pairs can be trapped by the QD, and quickly decay non-radiatively into the excited state of the QD forming an exciton state.  Radiative decay of this exciton leads to the emission of a photon. In practice, the QD can be excited to a higher excited state leading to a biexciton (2X) state (2 electrons, 2 holes), shown in Fig. \ref{fig:electronic_structure} (c) or to a higher multiexcitonic state (N electrons, M holes in QD).  Due to asymmetries in the QD, there is actually a fine structure splitting in the exciton state due to the different electron and hole spin states of the QD, which lifts the degeneracy of the exciton level due to the electron-hole exchange interaction, leading to slightly different transition frequencies for horizontally and vertically polarized light \cite{kulakovskii_fine_1999}.  The biexciton is a spin-singlet state which does not reveal a fine structure itself but decays to one of the two optically bright excitonic states. The polarization of the biexcitonic recombination lines is therefore also determined by the excitonic states.  An example of this splitting in the spectrum of single InP/InGaP QDs is shown in Fig. \ref{fig:electronic_structure} (d) \cite{ugur_single-photon_2012}. Using a polarizer
aligned at 0 and 90 degrees to the appropriate crystal axes, single lines at one of two different frequencies (corresponding to different spin states) are observed, as shown in the blue and red traces in Fig \ref{fig:electronic_structure} (d).  Removing or orienting this polarizer at 45 degrees allows the spectral lines corresponding to both spin states to be seen at the same time, as shown in the green trace in Fig \ref{fig:electronic_structure} (d). Higher order multi-excitonic and charged states can also be seen in the QD spectrum when the QD is pumped incoherently, as shown in the scheme in Fig. \ref{fig:electronic_structure} (b).  An example of such a spectrum is shown in Fig. \ref{fig:electronic_structure} (e) for an InAs QD in GaAs substrate, taken from reference \cite{santori_single-photon_2004}.  A single line must be spectrally isolated to obtain a single photon source; one way to isolate such a transition is using a high quality optical cavity \cite{gerard_strong_1999} (see section \ref{section:microcav}).  Under resonant excitation, only the exciton line can be seen, as shown in Fig. \ref{fig:electronic_structure} (f), also taken from reference \cite{santori_single-photon_2004}.  This will be discussed in more detail in section \ref{section:types_of_excitation}.

    \subsection{Charge states and properties}
For an odd number of particles in the QD, charged excitons are formed.  The simplest charged excitonic configuration is a trion (X$\pm$) and consists of one  exciton plus a single electron or hole. There is no fine structure splitting for a trion \cite{cade_fine_2006}, and the polarization of the emitted photon is determined by the spin of the excess carrier in the dot. Recombination lines from the trion state of the dot can be used for single photon sources \cite{ulrich_correlated_2005, strauf_high-frequency_2007}; advantages are the lack of fine-structure splitting and the lack of a dark state, the absence of which can lead to higher efficiencies, with a calculated increase in count rates of up to three times at high pump rates \cite{strauf_high-frequency_2007}. Charged excitons can also be used to create a $\Lambda$-system by applying a strong magnetic field orthogonally to the growth axis.  This will be discussed in more detail in section \ref{section:STIRAP}.

    \subsection{Growth}

Here we summarize the main mechanisms used for epitaxial QD growth.  The most common method of QD growth is self-assembly in Stranski-Krastanov (SK) mode.  Volmer-Weber (VW) growth occurs for larger lattice mismatches; fewer QD systems grown by this method have demonstrated single photon emission.  One reason for this is the greater size uniformity that can be obtained in the SK mode versus the VW mode, and the large strains involved in VW growth \cite{michler_single_2010}.

\subsubsection{Frank - van der Merwe (FM) growth}

In this mode, growth proceeds layer by layer, and it results in a very smooth epitaxial film \cite{frank_one-dimensional_1949}.  This mode can only occur when the lattice mismatch is not too high.  AlAs/GaAs growth system proceeds by FM growth.  This growth mechanism is used for growing the DBR structures and sacrificial layers used for optical microcavities (see section \ref{section:microcav}), in addition to the capping layers on QDs \cite{michler_quantum_2000}.

 \subsubsection{Stranski-Krastanov (SK) growth}

 In this mode the growth of a highly lattice-mismatched material initially proceeds layer by layer forming a planar wetting layer, until at some critical thickness of this layer self-assembled islands start forming. This occurs because the energy for island formation is lower than the strain energy to keep a planar wetting layer, which increases with the layer thickness. Most QD systems are grown in this mode using either molecular beam epitaxy (MBE) or metallo-organic chemical vapor deposition (MOCVD). MBE is most commonly used but MOCVD can also produce high quality low- and high-density quantum dots. InAs QDs on GaAs are generally grown by this method \cite{marzin_photoluminescence_1994, joyce_quantum_2005}.  An AFM image of uncapped InAs QDs grown by this method on GaAs substrate using MBE is shown in Fig. \ref{fig:qd_growth} (a).  The mechanism of growth is shown in the schematic below; a wetting layer forms followed by islands.

\subsubsection{Volmer-Weber (VW) growth}

In VW growth a large number of surface nuclei form initially, and then spread into 3D islands, unlike SK which occurs layer by layer until a critical thickness is reached \cite{volmer_zphys_1926}. Thus VW growth often results in a high mosaicity of the material inside the layer, and quantum dots grown by this method do not have a wetting layer.  It is a less common growth mode for QDs than SK, and occurs when the lattice mismatch is very high.  Examples of QDs grown by this method are InP/GaP QDs \cite{hatami_inp_2003} and InAs/GaP QDs \cite{leon_self-forming_1998}.

\subsubsection{Droplet epitaxy}

Droplet epitaxy is another different growth mechanism for quantum dots. In this mode group-III droplets are deposited on the substrate and then crystallized by exposing them to a group-V flux. This method can be used to grow both lattice mismatched (e.g. InAs/GaAs) and lattice matched (e.g. GaAs/AlGaAs) material systems. In contrast to the Stranski-Krastanov method there is no wetting layer. Droplet epitaxy is done at low temperature so annealing is usually necessary to improve the optical quality of the dots. The droplet epitaxy process is shown in Fig. \ref{fig:qd_growth} (b), reproduced from \cite{mano_fabrication_2000} for InAs QDs grown on 100 GaAs.  A droplet of In is deposited on the substrate, the flux of As crystalizes the edges of the droplets leading to a crater-like structure which is later annealed to form QDs.

\subsubsection{Site controlled quantum dots}
\begin{figure}
\includegraphics[width = 15cm]{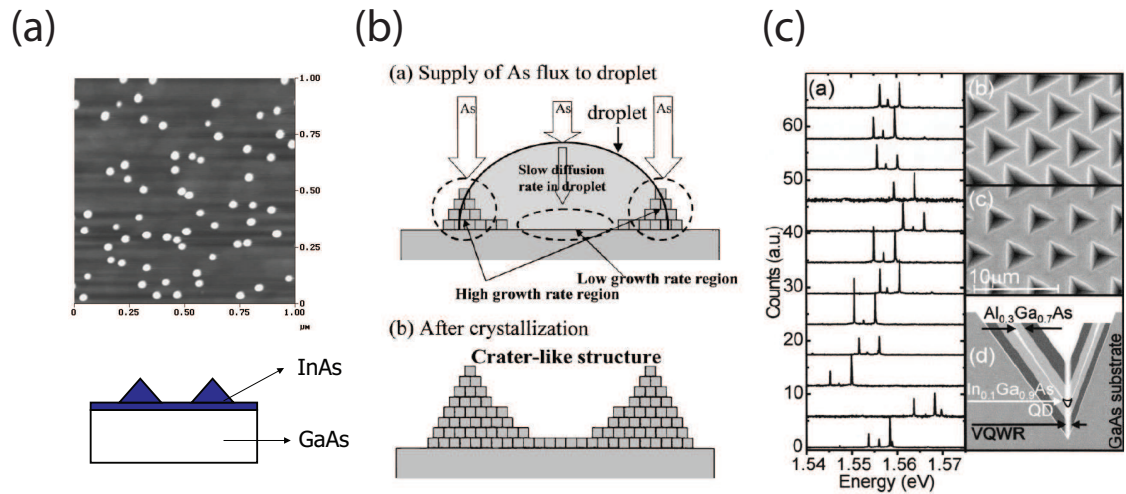}
\caption{(a)AFM image of uncapped SK grown InAs QDs on GaAs, and a schematic of how the QD islands form (AFM courtesy of Bingyang Zhang, Stanford). (b) Mechanism of droplet epitaxy QD formation. Reproduced with permission from ref. \cite{mano_fabrication_2000}. (c)(a) X$^{-}$ 2X and X emission peaks from a row of 12 different pyramidal QDs. Scanning electron micrograph of substrate before (b) and
after growth (c). (d) Schematic cross section of pyramidal QD structure. Reproduced with permission from ref. \cite{baier_high_2004}}
\label{fig:qd_growth}
\end{figure}

Self-assembled quantum dots have excellent properties as single photon emitters; however there is no control over their position and they have a broad inhomogeneous wavelength distribution.  This means that after choosing the correct density of quantum dots, integration with devices requires the fabrication of many structures to find one with a QD at the right wavelength coupled to it, or careful measurement and aligning before fabrication \cite{thon_strong_2009}.  Having control over the position and wavelength of the quantum dots is therefore highly desirable for scaling up quantum dot devices.  The first demonstration of single photon emission from a site-controlled quantum dot was in 2004 \cite{baier_single_2004} in the InGaAs/GaAs system.  These QDs are grown via lithographic positioning followed by etching of small holes in a predesigned pattern. QDs can be subsequently grown by MOCVD or MBE and will only grow in this holes.  Figure \ref{fig:qd_growth} (c) part (a) shows the spectral properties of 12 different site-controlled QDs grown in this way.  Parts (b) and (c) show the pyramidal etched holes, while (d) shows the cross section of a single InGaAs QD structure. These quantum dots were later integrated with photonic crystal structures \cite{gallo_integration_2008}.  These site controlled QDs are grown in 111-oriented GaAs material, and are an excellent solid-state source of polarization entangled photons \cite{mohan_polarization-entangled_2010}.  Site controlled quantum dots in the InP/GaInP system have also recently shown single photon emission in the red/visible wavelengths \cite{baumann_site-controlled_2012}, while work on site controlled quantum dots in the InGaAsN/GaAs system also shows promise \cite{carron_dilute-nitride_2011}.  Site controlled QDs emitting at 1550 nm have also been demonstrated \cite{song_site-controlled_2005}.

    \subsection{Materials systems and their properties}
    Various materials systems have been investigated as candidates for single quantum dot growth.  All of these have advantages and disadvantages in terms of their properties as single photon sources.  Wavelengths from the UV all the way to telecom wavelengths have been demonstrated, while site-controlled QDs have been demonstrated in many materials systems, and high and room temperature operation has also been achieved in wide band gap semiconductors. Here we attempt to list the main types of semiconductor quantum dots and their main features compared with each other.  Table 2 also shows a summary of this information.
        \subsubsection{InAs/GaAs}
        The most common quantum dots for single photon sources are InAs QDs on GaAs \cite{petroff_semiconductor_2011}.  These emit in the range of 850 nm - 1000 nm, and require cryogenic temperatures for operation due to the shallow carrier confinement.  These quantum dots are most usually grown in the Stranski-Krastanov mode by either MBE or MOCVD, although they can also be grown by droplet epitaxy.  First demonstrations of epitaxial QD single photon emission \cite{michler_quantum_2000}, quantum key distribution \cite{waks_secure_2002}, electrically pumped single photon emission \cite{yuan_electrically_2002}, integration with many different types of optical micro-cavities \cite{michler_quantum_2000, solomon_single-mode_2001, badolato_deterministic_2005}, strong coupling to optical micro-cavities \cite{reithmaier_strong_2004}, photon blockade \cite{faraon_coherent_2008}, resonant excitation \cite{flagg_resonantly_2009}, measurement of photon indistinguishability \cite{santori_indistinguishable_2002} and single photon laser \cite{nomura_laser_2010}, were all done with the InAs/GaAs quantum dot system. InAs/GaAs QDs can be capped with InGaAs to extend their emission wavelength to O-band telecommunications wavelengths, around 1300 nm at room temperature \cite{alloing_growth_2005, zinoni_time-resolved_2006}.  These capped quantum dots have been used for secure quantum key distribution over 35 km of fiber \cite{intallura_quantum_2007}. Different crystal orientations also lead to different quantum dot properties. Most QDs are grown on the (100) surface; however, more symmetric dots can be grown on (111) surfaces. The main advantage of this is that it translates into a minimal fine structure splitting. This is quite important for polarization entangled photon sources that use the biexciton-exciton radiative cascade \cite{benson_regulated_2000}. Due to the in-plane asymmetries (the dots are elongated in one direction) of conventional quantum dots, the exciton states are non-degenerate, separated in energy by the fine structure splitting. Since the two states are distinguishable in energy, this `which-path' information destroys the entanglement. This means that the more symmetric quantum dots grown on (111) surfaces are promising for entangled photon sources \cite{mano_self-assembly_2010, stock_single-photon_2010}.

        \subsubsection{III-P based emitters} The most efficient Si single photon detectors have maximum detection efficiency in the red part of the visible spectrum.  Quantum dots emitting in the red have been extensively studied over the past decade in materials systems including InP/InGaP,\cite{zwiller_generating_2003, ugur_single-dot_2008, luxmoore_control_2010, richter_low-density_2010, reischle_triggered_2010}, InP/GaP \cite{hatami_inp_2003, hatami_radiative_2001} GaInP/GaP, \cite{gerhard_structural_2009} InAs/GaP, \cite{leon_self-forming_1998}, AlGaInP/GaP \cite{gerhard_short_2011}, InGaAs/GaP \cite{rivoire_photoluminescence_2012} and  InP/AlGaInP, \cite{schulz_inp/algainp_2011, schulz_optical_2009}.   Due to the deep confining potentials, these QDs can work at higher temperatures than InAs/GaAs system.  Clear single quantum dots with narrow emission lines exhibiting anti-bunching have been observed only in the InP/InGaP \cite{ugur_single-photon_2012} and InP/AlGaInP systems, and an electrically pumped single photon source operating at up to 80K has been demonstrated in the InP/AlGaInP system \cite{reischle_electrically_2008}.  GaP in particular is an attractive material for QDs. It is almost lattice matched with Si; therefore GaP-based materials allow either monolithic integration with Si \cite{andre_heteroepitaxial_1975} or growth on a nonabsorbing GaP substrate (due to the large indirect electronic band gap) \cite{rivoire_photoluminescence_2012}. Additionally, the stronger second order optical nonlinearity of GaP compared to InGaP is preferable for on-chip frequency conversion to telecom wavelengths (see section \ref{section:freqconversion}).

        Recently, work on InAs/InP QD system has led to the development of C-band telecommunications wavelength QD single photon emitters.  Single photon emission from this system in the O-band range around 1300 nm was first observed in 2004 \cite{takemoto_non-classical_2004} by etching mesas in high density material.  Single photon emission at 1.55 $\mu$m was later observed in the same system \cite{miyazawa_single-photon_2005, cade_optical_2006, takemoto_optical_2007}.  This system has also shown electrically pumped single photon emission \cite{miyazawa_exciton_2008} and been coupled to photonic crystal cavities with Purcell enhancement of 11 \cite{birowosuto_fast_2012} with a lifetime reduction from 2.2 ns to 0.2 ns.

        \subsubsection{Wide band gap emitters}
For good reviews on wide band gap emitters see references \cite{bacher_optical_2003}, \cite{michler_single_2010} and \cite{arakawa_advances_2006}. Wide band gap QDs include (In,Ga)N QDs with (Ga,Al)N barriers \cite{arakawa_advances_2006, santori_photon_2005, kako_gallium_2006}, as well as self-organized CdTe \cite{terai_zero-dimensional_1998, tinjod_cdte/zn1xmgxte_2004, couteau_correlated_2004} and CdSe \cite{flack_near-field_1996, kummell_single_1998} QDs, which can be combined with barrier materials \cite{sebald_single-photon_2002, arians_room_2007, fedorych_room_2012}.  Large electronic band offsets are possible in these systems, which together with the small size of the dots leads to strong carrier confinement.  This allows higher temperature operation than in other systems. These also emit in the visible or even the ultraviolet spectral range. In quantum cryptography applications this could allow for reduced size in emitter/receiver telescopes.  Plastic fibers can also have transmission windows in this wavelength range. Very recently, single photon emission at room temperature has been demonstrated in CdSe/ZnSSe/MgS system \cite{fedorych_room_2012}, grown on a GaAs substrate, with $g^{(2)}(0)$ as low as 0.16, although at these high temperatures, linewidths become significantly broadened.  Electrically pumped single quantum dot emission has also been seen in both II-VI \cite{arians_electrically_2008} and nitride based systems \cite{kalden_electroluminescence_2010}, although as yet neither has demonstrated clear anti-bunching while electrically pumped.  Fabrication of optical microcavities in both systems is more challenging than in the III-As and III-P systems.  Cavity-enhanced single-photon emission from a single InGaN/GaN quantum dot has been demonstrated \cite{jarjour_cavity-enhanced_2007}, and similarly for II-VI systems \cite{lohmeyer_enhanced_2006, robin_purcell_2005}.

\begin{table}
  \centering
  \begin{tabular}{| l || l | p{1.2cm} | p{1.4cm} |p{3.5cm} |p{1.5cm} |}
  \hline
    Material System & $\lambda$ (nm) & $\tau$ (ns) & $T_{max} (K)$ & Comments & ref \\ \hline\hline
    InAs/GaAs & $\sim$850 - 1000 & $\sim$1& 50 &  & \cite{santori_single-photon_2004}\\ \hline
    InGaAs/InAs/GaAs & $\sim$1300 & $\sim$1.1-8.6  & 90 & biexponential decay & \cite{alloing_growth_2005, zinoni_time-resolved_2006}\\ \hline
    InP/InGaP & $\sim$650-750 & $\sim$1 & 50 &  & \cite{zwiller_generating_2003, ugur_single-dot_2008, luxmoore_control_2010, ugur_single-photon_2012} \\ \hline
    InP/AlGaInP & $\sim$650-750 & $\sim$0.5-1 & 80 &  & \cite{schulz_optical_2009} \\ \hline
    InAs/InP & 1550 & $\sim$1-2 & 50-70 &  & \cite{takemoto_observation_2004, cade_optical_2006, birowosuto_fast_2012}\\ \hline
    GaN/AlN & $\sim$250-500 & $\sim$0.1-1000 & 200 & lifetime increases with wavelength & \cite{santori_photon_2005, arakawa_advances_2006, kako_gallium_2006}\\\hline
    InGaN/GaN & $\sim$430& $\sim$8-60 & 150 &  & \cite{jarjour_cavity-enhanced_2007, kalden_electroluminescence_2010}\\ \hline
    CdTe/ZnTe & $\sim$500-550  & $\sim$0.2 & 50 &  & \cite{couteau_correlated_2004} \\ \hline
    CdSe/ZnSSe & $\sim$500-550 & $\sim$0.2 & 200 &  & \cite{sebald_single-photon_2002, lohmeyer_enhanced_2006}\\ \hline
    CdSe/ZnSSe/MgS & 500-550 & $\sim$1-2 & 300 & linewidths broaden significantly after 100 K & \cite{fedorych_room_2012}\\ \hline
    \end{tabular}
    \label{table:qd_properties}
\caption{Comparison of different epitaxially grown QD materials systems and their properties. $\tau$ is the lifetime of the QDs, $\lambda$ is the wavelength and $T_{max}$ is the maximum temperature at which single photon emission has been reported (although many of these systems still exhibit photoluminescence at higher temperature.}
\end{table}

    \subsection{Types of excitation}
    \label{section:types_of_excitation}
        \subsubsection{Continuous wave/pulsed}
One advantage of two-level emitter single photon sources over other processes such as spontaneous parametric downconversion (SPDC) is that the single photon source produces a single photon in response to an external trigger.  For optical excitation, this trigger is an optical pulse, and a single photon will be generated by each of these pulse triggers.  In the case of continuous wave (CW) excitation, the single photons are no longer triggered.  For single photon sources based on SPDC, these are often CW pumped, as they use one of a generated photon pair to herald its twin photon.  Protocols that can use heralded single photons and entangled photon pairs instead of triggered single photons for various QIP applications have been devised \cite{peters_towards_2006}.
        \subsubsection{Above band (non-resonant) optical}
        \label{section:above-band}
        Experimentally, it is very convenient to use above band excitation to excite QDs into the excited state.  Practically, it means that the low power single photon signal and high power pump laser can be easily separated spectrally, and no specific excitation wavelength is necessary.  The QD is excited above the band gap of the surrounding semiconductor (which for GaAs at low temperature is around 817 nm, a convenient wavelength for pumping with a pulsed Ti:Sapphire laser). Electron-hole pairs are mainly generated in the surrounding semiconductor.  Some fraction of these are captured by the wetting layer and fall into the excited states of the QDs where they quickly relax to the lowest energy levels via phonon assisted relaxation within a short time scale ($\sim$ 10-100 ps).  If the QD radiative recombination time is longer than the recombination time of the free electron-hole pairs in the semiconductor, each excitation pulse can lead to at most one photon emission event at the corresponding excitonic transition. Even when a single QD is isolated, several spectral lines are typically seen in photoluminescence. These exciton lines are at different frequencies and can be spectrally filtered to give single photon emission.  Loss of indistinguishability occurs when the phonon assisted relaxation process of carriers captured in the QD is not short compared with the QD radiative lifetime.  This adds an additional delay  to generated photons due to relaxation jitter \cite{santori_indistinguishable_2002}.  For more details see section \ref{section:qdindistinguishability}.

        \subsubsection{Quasi-resonant optical}

Quasi-resonant optical excitation involves exciting the quantum dot on transition with a higher excited state, e.g. the p-shell.  These higher excited states have broad linewidths due to their rapid relaxation.  A large laser power may be required since the absorption cross section of a single QD is small.  In this scheme it is possible to controllably inject a single electron-hole pair in the p-shell \cite{toda_efficient_1999, kamada_exciton_2001}. After relaxation into the first excited state (s-shell), a single photon can be emitted and a high quantum efficiency is possible. Another important aspect is a suppressed multi-photon emission giving a lower $g^{(2)}(0)$.  Dephasing processes should be drastically reduced since the charge carriers are exclusively generated within the desired dot; in the case of Eq. \ref{eq:indistinguishability} the relaxation rate, $\delta$, from higher order excited states should be faster, leading to a faster and more indistinguishable single photon source. Off-resonant coupling between a QD and a cavity via phonons is another quasiresonant excitation method.  In this case the cavity resonance and QD transition have a large spectral detuning, and non-resonant transfer of energy occurs via phonon-induced processes.  More detailed discussions of this phenomenon can be found in \cite{hennessy_quantum_2007,ates_non-resonant_2009,winger_explanation_2009,englund_resonant_2010,majumdar_linewidth_2010}.

\subsubsection{Resonant optical}
\begin{figure}[h!]
\includegraphics[width = 15cm]{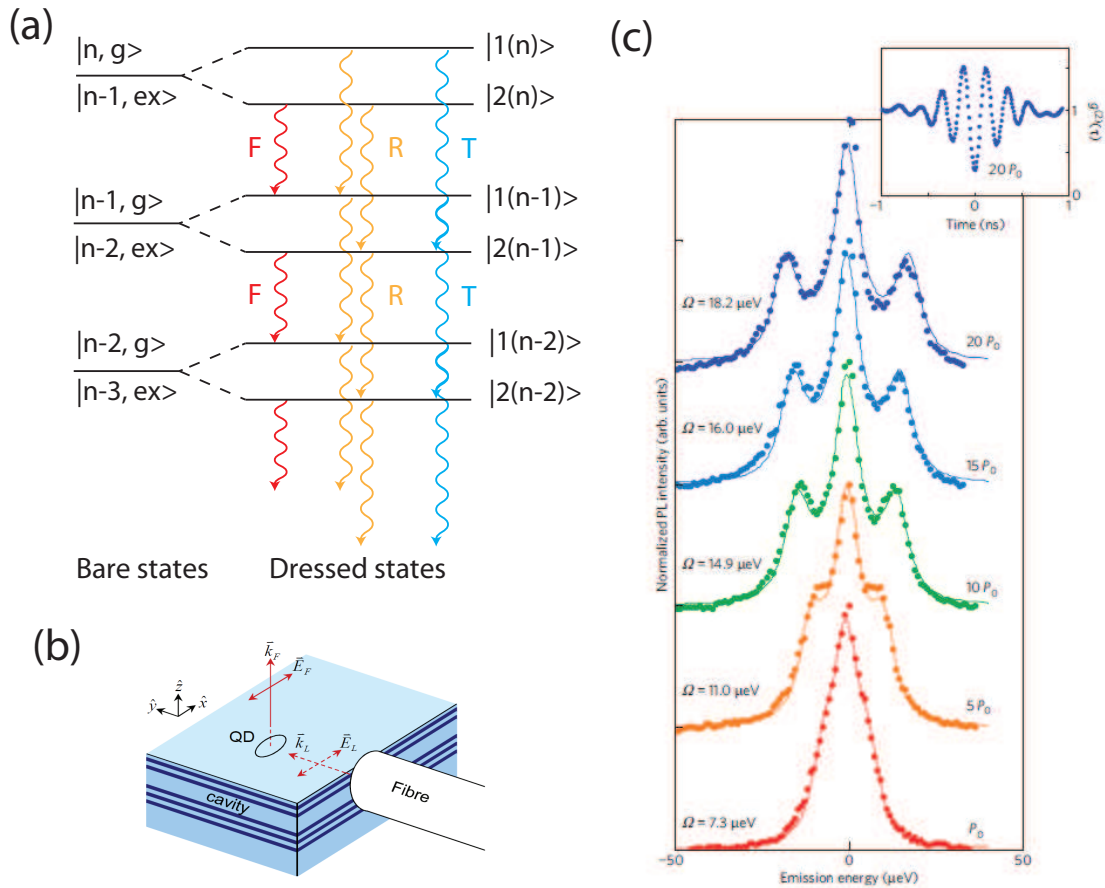}
\caption{(a) Dressing of the bare cavity and light field states. The transitions between these dressed states give a three peaked spectrum, known as the Mollow triplet. The three lines of the Mollow triplet are the Rayleigh (R) central line, with a low energy fluorescence sideband (F) and a higher energy three photon (T) sideband, corresponding to the transitions shown. (b) The experimental geometry for resonant excitation scheme. Laser light is supressed via distributed Bragg reflection in the cavity. (c) Normalized emission spectra from a QD at different excitation powers, distinctly resolving the Mollow triplet. The lines are fits and the Rabi energies are noted on the plot. The inset shows that the QD also shows oscillatory g$^{(2)}(\tau)$, although g$^{(2)}(\tau) \neq 0$. (b) and (c) reproduced with permission from \cite{flagg_resonantly_2009}}
\label{fig:resonant_excitation_fig}
\end{figure}

Resonant excitation into the first excited state (s-shell) of a QD is the most desirable form of excitation, as no additional relaxation process from a higher excited state is necessary before the photon is emitted (i.e. $\delta = \infty$ in Eq. \ref{eq:indistinguishability}), giving the highest indistinguishability of these processes.  This is difficult to implement practically, as it is challenging to separate the strong excitation laser pulse from a generated single photon.  Theoretically \cite{mollow_power_1969,cohen-tannoudji_atom-photon_1998}, for a Rabi frequency greater than the spontaneous emission rate, resonant optical excitation should produce a fluorescence spectrum with three peaks.  This occurs because the bare states of the two-level system are dressed by the strong interaction with the laser field.  For zero detuning of the laser from the atomic transition, these bare states consist of degenerate levels: ground state of emitter and n-1 pump photons ($|g, n-1\rangle$), excited state and n-2 pump photons ($|e, n-2\rangle$) and ground state n pump photons ($|g, n\rangle$), excited state n-1 pump photons ($|e, n-1\rangle$), and so on.  The dressing of these states by a strong laser splits these degeneracies, forming a ladder of dressed states, depicted in Fig.
\ref{fig:resonant_excitation_fig} (a). This can be derived by the same method as used for the strong coupling regime of
 an atom-cavity system which will be described in section \ref{section:strongcoupling}, and the situation in the case of photon blockade, described in section \ref{section:photonblockade}.  However, in this case the photon number is very large (n $\rightarrow\infty$), and the `cavity volume' $V \rightarrow\infty$  \cite{cohen-tannoudji_atom-photon_1998}.  The fluctuations in photon number go as $\sqrt{n}$ and in this case of large photon number can be neglected (see ref. \cite{cohen-tannoudji_atom-photon_1998} for details). Additionally, the spontaneous emission rate of the atom in this case remains unchanged.  Due to the large photon number, the spacing between rungs on the ladder is equal, unlike in the case of photon blockade.   Four optical transitions are allowed between these states; two of these transitions are degenerate. This gives a three peaked spectrum, known as the Mollow triplet.   The three lines of the Mollow triplet are the Rayleigh (R) central line, with a low energy fluorescence sideband (F) and a higher energy three photon (T) sideband (Fig. \ref{fig:resonant_excitation_fig} (a)).

Initial resonant excitation experiments did not resolve the Mollow triplet and involved measuring a photocurrent or change in transmission induced when scanning a laser frequency over the QD ground to excited state transition. Recent experimental progress has allowed collection of resonantly excited single photons.  In one experiment, strong polarization and spectral filtering allowed the sidebands of the Mollow triplet to be observed for a charged QD \cite{vamivakas_spin-resolved_2009}, while a second recent experiment used an engineered waveguide coupled cavity DBR structure for exciting and collecting via different channels to suppress scattered laser light  \cite{flagg_resonantly_2009}.  This setup is shown in Fig. \ref{fig:resonant_excitation_fig} (b), and the resolved Mollow triplet is shown in Fig. \ref{fig:resonant_excitation_fig} (c).  A review article on these two experiments can be found in ref. \cite{santori_quantum_2009}.  The indistinguishability of such resonantly excited photons has been measured, with a post-selected visibility of 0.9 \cite{ates_post-selected_2009}. Intensity autocorrelation ($g^{(2)}$) measurements of the filtered emission from the F or T sidebands show
antibunching as the state of the emitter changes, while cross-correlation measurements between the F and T lines show photon bunching, indicating time-cascaded emission from the two lines \cite{ulhaq_cascaded_2012}.  Emission from the filtered R line shows Poissonian statistics (the emitter state doesn't change).  This means that the sidebands of the Mollow triplet can also be used as a single photon source \cite{ulhaq_cascaded_2012}. This single photon source can be frequency-tuned by over 15 times its linewidth via laser detuning; detuning the laser from the resonance increases the total Rabi splitting between the sidebands and the central peak.  The dephasing of
QDs excited in the Mollow regime has also been experimentally characterized \cite{ulrich_dephasing_2011}.

In the case where the Rabi frequency is less than the spontaneous emission rate, the Heitler regime \cite{heitler_quantum_1954},
the spectrum and coherence properties of the laser are imprinted on the resonance fluorescence photons.  The QD then generates single photons with laser-like coherence, free from the dephasing processes affecting the QD emission. This has been demonstrated
\cite{nguyen_ultra-coherent_2011, matthiesen_subnatural_2012}. Resonant electrical injection via Coulomb blockade and electron and hole tunneling have also been proposed \cite{imamoglu_turnstile_1994} and demonstrated \cite{kim_single-photon_1999}. Finally, the Mollow triplet of a QD has been probed by combining resonant excitation of a single QD and collection from an off resonant cavity via phonon assisted interaction \cite{majumdar_probing_2011}. This approach provides a simpler experimental configuration, as excitation and output are spectrally separated.

        \subsubsection{Electrical}

        Electrical injection of a QD can be performed by growing the dot within a p-i-n junction.  Applying a short electrical pulse allows electrons and holes to cross the tops of the barriers and into the QD.  Most electrical injection schemes lead to the same indistinguishability problems as incoherent pumping, although using Coulomb blockade for resonant electrical injection has been proposed \cite{imamoglu_turnstile_1994} and demonstrated \cite{kim_single-photon_1999}. An overview of the research done on electrically pumped QD single photon sources will be given in section \ref{section:electrical}.

        \subsection{QD performance as a single photon source}

        \subsubsection{Measurement of $g^2(\tau)$}
\label{section:measurementofg20}
     \begin{figure}
\includegraphics[width = 15cm]{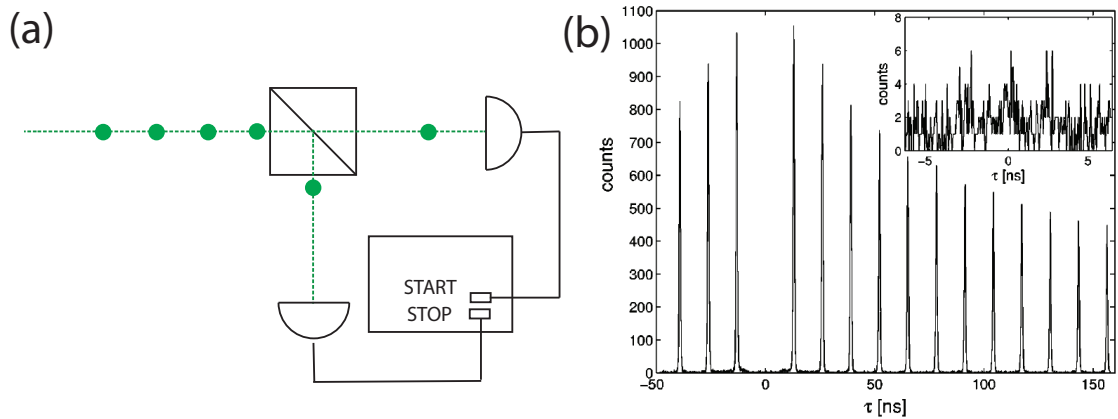}
\caption{(a)HBT measurement setup.  (b) Pulsed $g^{(2)}(\tau)$ measurement.  The missing central peak indicates single photon emission.  Inset shows missing signal at $\tau = 0$.  Reproduced with permission from ref. \cite{vuckovic_enhanced_2003}.}
\label{Fig:isogyre}
\end{figure}

The next problem is how to measure $g^2(0)$ experimentally.  In principle, a perfect detector with perfect time resolution could measure the times of single-photon events and calculate the correlation function directly. Although this has recently been demonstrated experimentally for the first time \cite{steudle_quantum_2011}, the detectors most commonly used for these measurements typically cannot perform such a measurement, due to dead times on the order of 1 ns. This means that after detecting the presence of a single photon, the detector cannot again measure for 1 ns.  To overcome this problem, detection schemes using two independent detectors in a Hanbury Brown and Twiss (HBT) \cite{brown_correlation_1956, scully_quantum_1997} type setup are usually used. In this setup, the single photons are sent to a 50/50 beam splitter, which with equal probability will send photons to one or the other of two single photon detectors.  Present state of the art detectors are avalanche photodiodes (APDs), which offer detection efficiencies $\sim$ 40-70 \% in the visible and near infrared spectrum and have response times of 400-700 ps. Lower efficiency APDs ($\sim$ 5-35 \%) with faster response times (30-50 ps range) are also available. For an up-to-date review of available single photon detectors see the review article by Eisaman et al. \cite{eisaman_invited_2011}. In the most commonly used detection mode, only the time differences $\tau$ between the detection events (start and stop) are registered and in a subsequent process a time-to-amplitude conversion followed by a multichannel analysis is performed in order to generate a histogram of coincidence events $n(\tau)$. The measured coincidence function $n(\tau)$ differs from the original second order coherence function $g^{(2)}(\tau)$.  The probability to measure a time difference at time $\tau$ is given by \cite{michler_single_2010}: $n(\tau)$ = (probability to measure a stop event at time $\tau$ after a start event at time 0) $\times$ (probability that no stop detection has occurred before)
\begin{equation}
n(\tau) = (G^{(2)}(\tau)+R_{dark})(1-\int^{\tau}_0n(\tau')\mathrm(d)\tau'),
\end{equation}
where $G^2(\tau)$ is the unnormalized second-order coherence function and $R_{dark}$ describes the detector dark counts \cite{michler_single_2010}. The measured histogram of coincidence counts $n(\tau)$ approaches $G^{(2)}(\tau)$ in the limit when $R_{dark}$ is much smaller than the signal count rate $R$, and the average arrival time of the photons 1/R is much smaller than the observed delay time $\tau$. This means that the probability that no stop detection has occurred before, is approximately 1. Losses, like undetected photons lead only to a global decrease of $G^{(2)}(\tau)$ which can be compensated, e.g., with a longer measuring time.

In practice, long measurement times are often necessary.  The response and dead time of the counters means that if two events occur too close in time to each other, they cannot be resolved.  Experimentally, the detection count rate should stay below this rate in order to avoid this error, and the collection time can be correspondingly increased to build up the histogram.  For very low count rates (e.g. due to poor quantum efficiency or poor collection efficiency), very long collection times are necessary, in which case sample drift can become an issue, and active stabilization may be necessary.

\subsubsection{QD single photon source indistinguishability}
\label{section:qdindistinguishability}
The excitation of a single QD is a rapid process compared to the subsequent spontaneous decay back to the ground state.  Therefore a bare QD single photon source speed is limited by the spontaneous emission lifetime of the quantum dot, which is in the nanosecond regime.  A major drawback of the nonresonant excitation process is that charge carriers can be captured by adjacent traps or defect centers in the vicinity of the dot. This might lead to fluctuations in the emission wavelength between different pulses and is known as spectral diffusion, a major line broadening effect for quantum dot transitions.  Two emitted photons separated by a time interval longer than the spectral diffusion time will be distinguishable in principle, because their
frequencies will differ and they will not interfere \cite{santori_indistinguishable_2002}. However, if the delay between the emission times of the two photons is short enough, slow spectral diffusion processes may be neglected.  An additional loss of indistinguishability arise from above-band excitation, discussed in section \ref{section:above-band}.  This causes time jitter that affects the temporal
overlap of the single photon pulses. The indistinguishability in the case of above-band excitation is given by
        \begin{equation}
        \label{eq:indistinguishability}
        I = \frac{\Gamma}{\Gamma+\alpha}\frac{\delta}{2\Gamma+\delta}
        \end{equation}
        where $\alpha$ is the phonon dephasing rate of the excited state and $\delta$ is the relaxation rate from the higher-order excited states to the first excited state (from which the single-photon pulse is emitted), leading to a jitter in the arrival time of the single photon wavepacket \cite{vuckovic_generation_2006}.  This expression leads to an optimal value for the radiative lifetime for maximizing $I$, obtained by differentiating the expression for $I$ with respect to $\Gamma$, and giving $\Gamma = \sqrt{\alpha \delta/2}$, which has a value for InAs QDs of around 100-140 ps \cite{santori_indistinguishable_2002}. This value can be achieved using microcavities to enhance the radiative emission rate (see section \ref{section:microcav}).  With the optimal $\Gamma$ and realistic values of $\alpha$ and $\delta$, the achievable I=70-80\% \cite{vuckovic_generation_2006}. For higher emission rates there is therefore a tradeoff between speed and indistinguishability.

\begin{figure}
\includegraphics[width = 15cm]{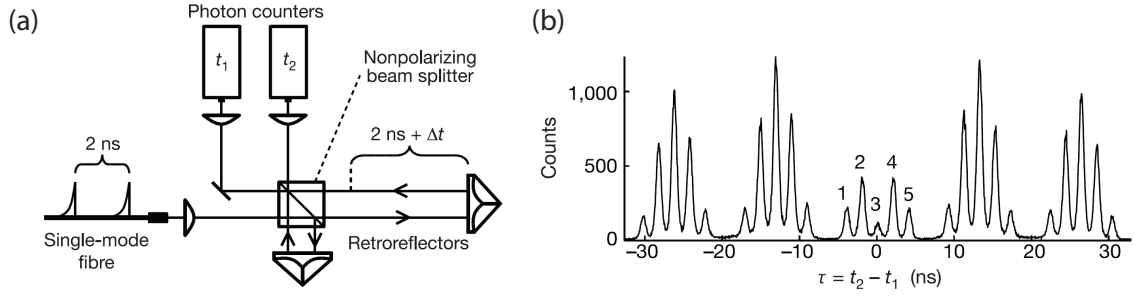}
\caption{Two-photon interference experiment. (a) Experimental setup. Single photon pulses separated by 2 ns are introduced every 13 ns through a single mode fiber. The pulses interfere in a michelson interferometer. Corner-cube retro-reflectors are used to increase tolerance to mode misalignment. (b) Histogram (53-ps bin size) obtained for a QD, with $\Delta t = 0$. The small area of peak 3 demonstrates two-photon interference. Figures reproduced with permission from ref. \cite{santori_indistinguishable_2002}.}
\label{fig:santori_indistinguishable_figure}
\end{figure}

The indistinguishability of photons from a single-photon source can be measured by colliding two individual photon wavepackets at a beam splitter in a Hong-Ou-Mandel-type experiment \cite{hong_measurement_1987}. The statistics of the outcome of the photons from the beam splitter is detected by single-photon detectors. If the duration of the single photon wavepackets exceeds the response time of the detectors, interference effects occur and can be studied in a time-resolved manner.  However, in order avoid measuring slow spectral diffusion, the indistinguishability can be measured on a shorter timescale.  In an experiment by Santori et al. \cite{santori_indistinguishable_2002}, an InAs/GaAs QD  was excited by a pair of laser pulses separated by $\Delta T$ ($\sim$ 2-3 ns) with a laser repetition period of $\sim$ 13 nm. The setup, reproduced from ref. \cite{santori_indistinguishable_2002} is shown in Fig. \ref{fig:santori_indistinguishable_figure} (a).  The QD will emit a single photon with each pulse. After polarization selection, these emitted photons are sent to the two arms of the Michelson interferometer (MI), which introduces a propagation delay between the short and long arms of $\Delta T+\delta t$. The two output ports of the beam splitter are fed to single photon counting modules where the time differences between the detection events (Start ($t_1$) and Stop ($t_2$)) are registered and a histogram of coincidence events of the time intervals $\tau=t_2-t_1$ is developed. Fig. \ref{fig:santori_indistinguishable_figure} (b) presents such a histogram for $\delta t = 0$. The histogram shows clusters of five peaks separated by the pump laser repetition period. The five different peaks are due to different combinations of photon paths through the interferometer. The peaks at $\tau = \pm2 \Delta T$ (1,5) arise from the first photon taking the short arm and the second taking the long arm. For the peaks at $\tau = \pm \Delta T$ (2,4) both photons pass through the same arm. The central peak $\tau = 0$ (3) corresponds to the situation where the first photon takes the long arm and the second photon takes the short arm causing both photons to arrive at the beam splitter at the same time. The reduced coincidence signal at $\tau = 0$ is the signature of the two-photon interference for this event. The probability of two photons colliding in the beam splitter and leaving in opposite directions can be defined by the quantity
\begin{equation}
p_{34}(\delta t) = \frac{A(0)}{A(T)+A(-T)},
\end{equation}
where A($\tau$) is the area of the peak at time interval $\delta t$ on the histogram where the delay of the MI interferometer is set to $\Delta T + \delta t$. The coincidence dip $p_{34}(\delta t)$ is then measured by varying the interferometer path length offset $\delta t$.

\section{Microcavity single photon sources}
\label{section:microcav}
Coupling a single emitter to a cavity is very desirable for a number of reasons.  These include higher repetition rates, high quantum efficiencies, and increased indistinguishability of emitted photons, all of which will be explained in more detail in the following section. A quantum dot will randomly emit single photons in any direction.  Coupling to a cavity will direct this emission into the cavity mode, which can be engineered to be easily coupled to fiber or to free space optics.  In addition, this cavity mode will have a well defined polarization, which is important for some linear optical quantum computing schemes. Examples of modern semiconductor cavity structures are micropillar cavities, microdisk cavities and photonic crystal cavities, \cite{vahala_optical_2003} which will be described in more detail in this section. These resonator structures are characterized by well defined spectral and spatial mode profiles as a consequence of a strong lateral and vertical confinement of the light. This confinement leads to very high quality factors in very small mode volumes.

 \subsection{Strong coupling regime}
 \label{section:strongcoupling}
 \begin{figure}
\includegraphics[width = 15cm]{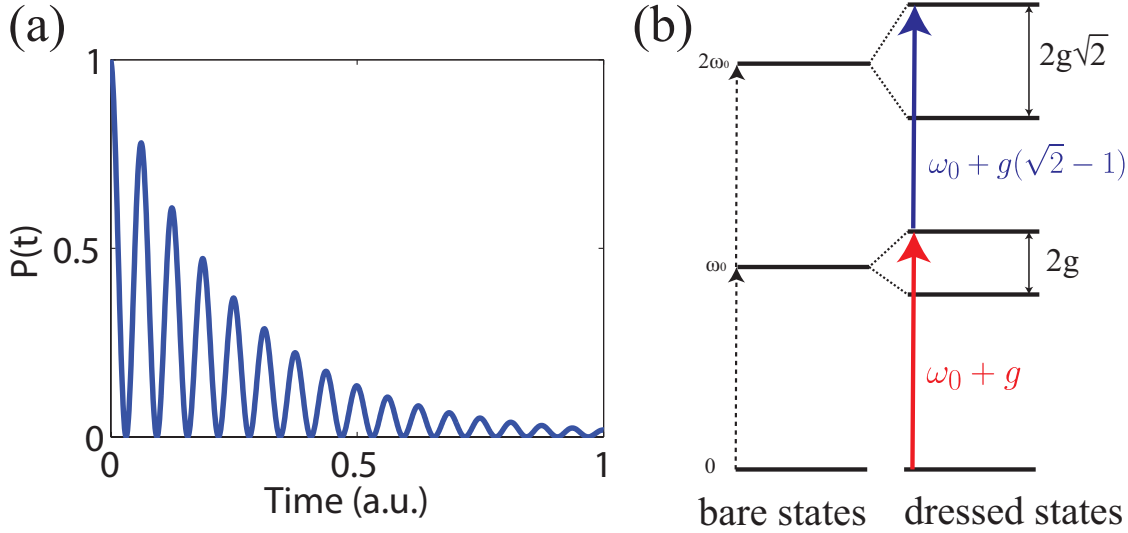}
\caption{(a) Rabi oscillation between the $|e, n\rangle$ state and $|g, n+1\rangle$ states. P(t) is the probability for the emitter to be in the $|e, n\rangle$ state, when $P(t) = 1$ the emitter is in the $|e, n\rangle$ state, and when $P(t) = 0$ the emitter is in the $|g, n+1\rangle$ state. (b) Jaynes Cummings ladder of states.}
\label{fig:strongcoupling}
\end{figure}

 Depending on the properties of the particular emitter-cavity system, the coupling of the cavity light field to the emitter will enter different regimes, displaying different characteristics.   In the strong coupling regime, the time scale of coherent coupling between the atom and the cavity field is shorter than that of the irreversible decay into various radiative and noradiative routes. Rabi oscillation occurs, and the time evolution of the system can be described by oscillation at frequency $2\sqrt{n+1}|g(\vec{r_A})|$ between the states $|e,n\rangle$ and $|g,n+1\rangle$, where $|e,n\rangle$ corresponds to an atom in the excited state and n photons in the cavity (i.e., the initial state of the system), and $|g,n+1\rangle$ corresponds to an atom in the ground state and n+1 photons in the cavity.  Such Rabi oscillations are illustrated in Fig. \ref{fig:strongcoupling} (a).  $g(\vec{r_A})$ is the coupling parameter between the cavity and emitter, given by

    \begin{equation}
    \label{equation:g}
    g(\vec{r}_A) = \frac{|\vec{\mu}_{eg}|}{\hbar}\sqrt{\frac{\hbar \omega}{2 \epsilon_M V_{mode}}}\psi(\vec{r}_A)cos(\xi),
    \end{equation}
    \begin{equation}
    \psi(\vec{r}_A) = \frac{E(\vec{r}_A)}{|E_{max}|}
    \end{equation}
    \begin{equation}
    \cos{\left(\xi\right)} = \frac{\vec{\mu}_{eg}\cdot{\hat{e}}}{|\vec{\mu}_{eg}|}
    \end{equation}

       where $\vec{\mu}_{eg}$ is the QD dipole moment, $V_{mode}$ is the cavity mode volume, $\epsilon_M$ is the material permittivity at the point of maximum $\epsilon|E|^2$ (where $E = E_{max}$) and $\vec{r}_A$ is the location of the emitter.  The value of $\psi$ gives the relative strength of the electric field at $\vec{r}_A$ compared to $E_{max}$, and $\cos({\xi})$ the fraction of the dipole moment along the direction of the electric field, $\hat{e}$ ($\vec{E} = E\cdot\hat{e}$).
 The condition for strong coupling depends on the strength of this coupling parameter, and is usually stated as
 \begin{equation}
 \label{equation:sc}
|g|>\kappa/2, \gamma,
 \end{equation}
 where $\kappa$ is the cavity field decay rate ($\kappa = \omega/2Q$) and $\gamma$ is the natural emitter decay rate (this can also be seen from the Eq. \ref{eq:omegapm} below, as in this regime the expression under square root becomes negative and two distinct solutions for real parts of the eigenfrequencies appear).  Looking at Eqs. \ref{equation:g} and \ref{equation:sc} we can see that in order to reach the strong coupling regime, it is necessary to increase the Q-factor and simultaneously reduce the cavity mode volume, place the atom (or exciton, in case of the solid-state cavity QED) at the location of the maximum field intensity and align the atomic (excitonic) dipole moment with the cavity field polarization.

Let's now explain the meaning and origin of this condition. The unperturbed Hamiltonian of the atom cavity system is given by
\begin{equation}
H_0 = H_A+H_F,
 \end{equation}
  where $H_A = \frac{\hbar \nu}{2}\hat{\sigma_z}$, $H_F = \hbar \omega \left(\hat{a}^{\dagger} \hat{a}+\frac{1}{2}\right)$ and $\hat{a}, \hat{a}^{\dagger}$ are the annihilation and creation operators for the light field, $\hat{\sigma}_{+}, \hat{\sigma}_{-}$ are the atom population operators. The unperturbed eigenstates, known as the bare states are given by $|e, n\rangle$ and $|g, n+1\rangle$, with eigenenergies $\hbar\omega(n+1/2)$.

 Once strongly coupled, this Hamiltonian must include a perturbation in the form of an atom-cavity interaction term, and the atom and cavity must be treated as a single system with an anharmonic ladder of states (Jaynes-Cummings model) \cite{scully_quantum_1997}. The Jaynes-Cummings Hamiltonian is
 \begin{equation}
 H = H_A+H_F+H_{int}
 \end{equation}
 and
 \begin{equation}
 H_{int} = i\hbar\left(g^{*}(\vec{r}_A)\hat{a}^\dagger\hat{\sigma}_{-}-g(\vec{r}_A)\hat{\sigma}_{+}\hat{a}\right).
 \end{equation}
 When the interaction Hamiltonian is turned on, the bare eigenstates are coupled (coupling to other states is neglected by the rotating wave approximation).  This coupling leads to the new eigenstates of the Hamiltonian H, $|1n\rangle$ and $|2n\rangle$, which are referred to as the dressed states, and have corresponding eigenenergies $\hbar(\omega \pm g\sqrt{(n+1)})$.  Therefore the dressed states are not degenerate, and exhibit a splitting $2\hbar|g|\sqrt{(n+1)}$, dependent on the photon number $n$.  This splitting is usually used as the indication that the emitter-cavity system has reached the strong coupling regime.   A ladder of dressed states is formed in the strong coupling regime, as is illustrated in Fig. \ref{fig:strongcoupling} (b).  This ladder is anharmonic, i.e. the splitting between dressed state energy levels is not constant.  This anharmonicity leads to effects such as photon blockade, which will be discussed in section \ref{section:photonblockade}.

 In the presence of detuning between the atom and the cavity, the two lowest order eigenstates have frequencies of $\omega_{\pm} = \hbar\omega\pm\sqrt{(\hbar\delta/2)^2+(\hbar g)^2}$, where $\delta = \nu-\omega$, and $\nu$ and $\omega$ are atom and cavity frequencies, respectively.  In the presence of losses, the resulting eigenfrequencies can be phenomenologically obtained by plugging in $\omega-i\kappa$ and $\nu-i\gamma$ into this expression, instead of $\omega$ and $\nu$ respectively. This leads to
 \begin{equation}
 \label{eq:omegapm}
 \omega_{\pm} = \frac{\omega+\nu}{2}-i\frac{\kappa+\gamma}{2}\pm\sqrt{\left(\frac{\delta-i(\kappa-\gamma)}{2}\right)^2+|g|^2}.
 \end{equation}
 As the system enters the strong coupling regime, for $|g|>>\kappa/2$ and $g>>\gamma$,
 \begin{equation}
 \label{equation:scdecay}
 \omega_{\pm}\rightarrow\frac{\omega+\nu}{2}\pm|g|-i\frac{\kappa+\gamma}{2}
 \end{equation}
Therefore, the eigenstates decay with the rate
\begin{equation}
\Gamma = (\gamma+\kappa)/2.
\label{eq:scrate}
\end{equation}
This is an upper limit on the decay rate of the emitter, and therefore the highest rate that the single photon source can achieve.

    \subsection{Weak coupling regime Purcell enhancement}

    In the weak-coupling case ($\gamma<g<\kappa/2$, corresponding to the `bad' cavity and narrow linewidth emitter), the irreversible decay rates dominate over the coherent coupling rate; in other words, the atom-cavity field system does not have enough time to couple coherently before dissipation occurs. This irreversible spontaneous emission process can be viewed as the result of an atom interacting with a large number of modes, and its attempt to start Rabi oscillations at different frequencies; this leads to destructive interference of probability amplitudes corresponding to different modes and to irreversible spontaneous emission.

    In this (Purcell) regime, the decay rate of the emitter can also be obtained from Eq. \ref{eq:omegapm} with $\kappa>>g>>\gamma$, and is equal to $g^2/\kappa$.
    Multiplying by 2 to give the energy decay rate gives a spontaneous emission rate
    \begin{equation}
     \Gamma = 2 \frac{|g(\vec{r}_A)|^2}{\kappa} = 2 \hbar |\vec{\mu}_{eg}|^2 \frac{Q}{\epsilon_M V_{mode}}\psi^2(\vec{r}_A)\cos^2(\xi).
       \end{equation}
    For a misaligned emitter, $\Gamma$ follows the same $\cos{}^2(\xi)|\psi|^2$ dependence as $g^2$ (see Eq. \ref{equation:g}). Clearly, $\Gamma$ can be  increased by increasing $Q/V_{mode}$ of the cavity. Off resonance with the cavity, the spontaneous emission rate follows a Lorentizian lineshape given by the cavity density of states, and the full expression for the modified spontaneous emission rate including detuning is given by
    \begin{equation}
 \Gamma = 2 \hbar |\vec{\mu}_{eg}|^2 \frac{Q}{\epsilon_M V_{mode}}\psi^2(\vec{r}_A)\cos^2(\xi)\cdot\frac{\Delta \lambda_c^2}{4\left(\lambda - \lambda_c \right)^2 + \left( \Delta\lambda_c \right)^2}
    \end{equation}
        where $\lambda_c$ is the cavity resonance wavelength and $\Delta\lambda_c =\lambda_c/Q$  is the cavity linewidth. The Purcell factor $F$ is the ratio of the modified spontaneous emission rate to the bulk emission rate of the emitter.  For a 3D photonic crystal cavity, the density of states far from the cavity resonance will be zero (see section \ref{section:PCC}).  For other structures, far off resonance from the cavity the spontaneous emission rate will be additionally modified by coupling to leaky modes.  Purcell enhancement was first demonstrated for a QD-cavity system in 1998 \cite{gerard_enhanced_1998}, with extensions soon after to single QDs \cite{gerard_strong_1999, michler_quantum_2000, santori_triggered_2001}.

   Purcell enhancement can also be used to increase indistinguishability, to match the spontaneous emission rate enhancement to the optimum value for high indistinguishability, as explained in section \ref{section:qdindistinguishability}. The maximum attainable spontaneous emission rate enhancement in the weak coupling regime occurs at the onset of strong coupling, and once strong coupling is achieved this
   speed is fixed. The advantage of operating a single photon source in the strong coupling regime is that the efficiency of a strongly coupled single
   photon source is close to one.  Moreover, different schemes in the strong coupling regime (such as photon blockade) can be employed to generate single photons with 100\%  indistinguishability  \cite{faraon_generation_2010}.  The atom-cavity system will also now have a much larger cross section than for a single atom. Section
   \ref{section:photonblockade} describes such a single photon source based on a strongly coupled atom-cavity system.  Additionally, the
    ideal single photon source for quantum information processing would act as both a single photon source and receiver, and could act as a
     node in a quantum network. For this ideal source, the single photon emission process must be reversible. This is not true in the case of incoherent above-band pumping, where the spontaneous emission process is irreversible and cannot be described by a Hamiltonian evolution. One alternative to incoherent pumping based on a strongly coupled atom-cavity system is stimulated Raman adiabatic passage (STIRAP); this is described in more detail in section \ref{section:STIRAP}.

        \subsection{Whispering gallery resonators}
Whispering gallery resonators rely on the confinement of light by total internal reflection at a curved boundary between two materials with different refractive indices. This results in the propagation of high-Q modes close to the boundary. Microdisk resonators are formed by etching disk-like shapes in semiconductor materials (usually Si, or III-V semiconductors, such as GaAs, or InP), and then partially wet etching underneath leaving a disk supported by a small post at the center. Such structures can support very high quality whispering gallery modes. Since the modes are mainly localized in the region close to the disk boundary, the presence of a small post supporting the disk at its center does not perturb the mode quality factor and volume significantly.  The maximum measured Q-factors for GaAs microdisks are around $\sim 10^5$ \cite{srinivasan_linear_2007}; the corresponding calculated mode volume is around 6($\lambda/n^3$), where the refractive index of the disk is $n$ = 3.6.  Strong coupling was first observed in this system in 2005 \cite{peter_exciton-photon_2005}.  Purcell enhancements in the weak coupling regime of around 8 have also been measured experimentally \cite{michler_quantum_2000, kiraz_cavity-quantum_2001}.  An SEM reproduced from \cite{michler_quantum_2000} is shown in Fig. \ref{fig:microcavities} (a).  Further Purcell enhancement is not beneficial for QIP schemes due to loss of indistinguishability, however for QKD and other applications where indistinguishability is not critical further enhancement could be helpful.

    \subsection{Micropost resonators}
     A micropost microcavity is formed by sandwiching a spacer (defect) region between two dielectric mirrors. Dielectric mirrors are distributed-Bragg-reflectors (DBR's), constructed by growing quarter-wavelength thick high- and low-refractive-index layers on top of each other. In the InAs/GaAs system, these are usually alternating layers of GaAs and AlAs  (corresponding to refractive index contrast of 3.6/2.9, respectively), and with GaAs as the spacer layer.  When the mirrors are infinitely wide in the lateral directions (directions perpendicular to the direction of DBR), the cavity is called a planar DBR cavity and is equivalent to a Fabry-Perot resonator. For both large Purcell enhancement and strong coupling, a small mode volume is as crucial as a large Q-factor for the majority of applications; for this reason, DBR structures are made with finite diameters. Such cavities are also referred to as DBR micropost microcavities.  Confinement of light in the structures with finite diameter is achieved by the combined action of the distributed Bragg reflection (DBR) in the longitudinal direction (along the post axis), and the total internal reflection (TIR) in the transverse direction (along the post cross-section). The spacer region is constructed by increasing the thickness of a single high- refractive-index region. Depending on the thickness of the spacer region and its refractive index, the localized mode can either have a node or an antinode of its electric field in the center of the spacer.  Microposts are usually rotationally symmetric around the vertical axis, although structures with exotic cross-sections, such as elliptical, square, rectangular have also been studied. For a micropost with rotational symmetry, DBR mirrors can be viewed as one dimensional (1D) photonic crystals generated by stacking high- and low-refractive-index disks on top of each other, and the microcavity is formed by introducing a defect into this periodic structure. The design of these structures for single photon source application requires that the QD be located at the field maximum, meaning that the spacer is designed to have the field maximum at the center of the spacer, where the QD will be grown. The Q factor can be increased by increasing the diameter of the cavity, but this will increase the mode volume of the cavity.  Therefore the design should be optimized depending on the application \cite{vuckovic_optimization_2002, pelton_three-dimensionally_2002}.  The first single photon source consisting of a single QD in micro-pillar cavity was demonstrated in 2001 \cite{moreau_single-mode_2001}.  Initial single photon sources based on this system showed efficient outcoupling and Purcell enhancement \cite{solomon_single-mode_2001, pelton_efficient_2002, vuckovic_enhanced_2003, santori_single-photon_2004}. For small diameter microposts factors of over 20,000 and cavities with theoretical Purcell enhancements of $>$ 75 \cite{daraei_control_2006} have been demonstrated, although as with microdisks experimentally measured enhancements have been considerably less than this \cite{vuckovic_enhanced_2003}.  For larger diameter microposts, quality factors of over 200,000 have been demonstrated \cite{reitzenstein_alas/gaas_2007}. Strong coupling was first demonstrated in a QD-micropillar cavity in 2004 \cite{reithmaier_strong_2004}.  The fastest quantum dot single photon sources yet demonstrated were in DBR cavities with speeds of up to 1 GHz \cite{bennett_electrical_2005, ellis_cavity-enhanced_2008}, as these are simpler to pump electrically and so are not limited by the laser modulation speed. An SEM and simulated electric field for a micropost resonator used as a single photon source is shown in Fig. \ref{fig:microcavities} (b), reproduced from \cite{vuckovic_enhanced_2003}.

    \subsection{Photonic crystal cavities}
    \label{section:PCC}
`Photonic crystals' refer to structures with periodic dielectric constants.  Some one-dimensional periodic media, such as the structures used in VCSELs, are instead referred to as Bragg reflectors, although the mechanism is the same. Other planar 1D periodic structures such as nanobeam \cite{foresi_photonic-bandgap_1997, deotare_high_2009} cavities are still referred to as photonic crystals.

\begin{figure}
\includegraphics[width = 15cm]{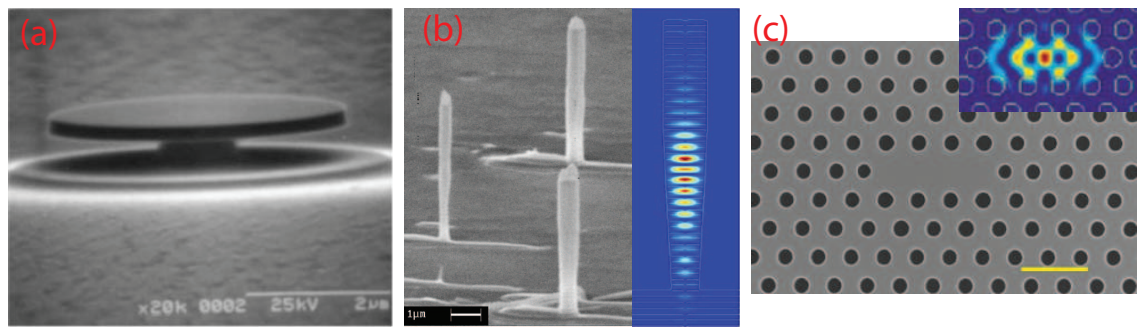}
\caption{(a) Microdisk cavity. Reproduced with permission from \cite{michler_quantum_2000}.  (b) Micropillar cavity and FDTD simulated electric field. Reproduced with permission from \cite{vuckovic_enhanced_2003}.  (c) Photonic crystal L3 cavity and simulated $|E|$ field.  The scale bar is 2 $\mu$m. Reproduced with permission from \cite{rivoire_second_2009}}
\label{fig:microcavities}
\end{figure}

 Three-dimensional (3D) photonic crystals can lead to a complete photonic band gap, meaning that in a certain frequency region, the wave propagation is prohibited through the crystal in any direction in space and for any polarization.  3D photonic crystals offer the opportunity for light manipulation in all three dimensions in space. They are very difficult to fabricate, although high Q 3D photonic crystal cavities with coupled QDs have been demonstrated \cite{tandaechanurat_lasing_2011}. For this reason, most of the research efforts in the past have been focused on planar photonic crystals, i.e., two dimensional (2D) photonic crystals of finite depth, which can be made by standard microfabrication methods. The light confinement in planar photonic crystals results from the combined action of distributed Bragg reflection in the 2D photonic crystal and internal reflection in the remaining dimension. The imperfect confinement in the third dimension produces some unwanted out-of-plane loss (radiation loss), which is usually a limiting factor in performance of these structures; still, most of the functionality of 3D photonic crystals can be achieved by careful design, exhibiting a photonic band gap for electromagnetic waves propagating in the plane of the crystal. For a 3D photonic crystal, inside the photonic band gap the density of optical states is zero, while outside the band gap, normal modes exist that can be classified based on their K-vector. For a planar 2D photonic crystal the density of optical states will not quite drop to zero inside the band gap, but will be greatly reduced relative to free space.  However, by perturbing a photonic crystal lattice (i.e., by introducing lattice defects), one can permit localized modes that have frequencies within the photonic band gap. Such modes have to be evanescent inside the photonic crystal, i.e., they have to decay exponentially away from the defect. In other words, the defect behaves as a cavity, and the surrounding photonic crystal represents mirrors surrounding the cavity. Therefore, the defects introduce peaks into the density of optical states inside the photonic band gap. Moreover, by analogy with localization of electron wavefunctions near impurities in semiconductors, the defects break the discrete translational symmetry of the photonic crystal, and one can no longer classify the modes based on their K-vector. The simplest way of forming a microcavity starting from the unperturbed hexagonal photonic crystal lattice of air holes, for example, is by changing the radius of a single
hole, or by changing its refractive index. The former method is more interesting from the perspective of fabrication, since lithographic tuning of parameters of individual holes is a simple process to implement. The highest Q-factors of photonic crystal cavities have been demonstrated in silicon at 1550 nm, where material absorption is minimal. Q-factors as high as $3 \times 10^6$ in Si have been reported recently  by passive, transmission measurements \cite{taguchi_statistical_2011}, while Q factors as high as 700,000 have been measured in GaAs \cite{combri_GaAs_2008}, although at 1.55 $\mu$m. At the shorter wavelengths of InAs QD emitters, operation closer to the GaAs band edge degrades Q.  The presence of QDs also increases absorption and lowers Q \cite{stumpf_light-emission_2007}.  Single photon sources based on photonic crystal cavities \cite{englund_controlling_2005} with deterministic positioning \cite{badolato_deterministic_2005} have been demonstrated.  These show Purcell enhancement of $>$10 \cite{toishi_high-brightness_2009}, with theoretical maxima of over 100 (if the QD was optimally aligned with the electric field maximum).  On chip transport of single photons generated on chip via waveguides is also possible \cite{englund_generation_2007}.  Since the Purcell enhancement depends on the quality factor of the photonic crystal cavity, dynamic tuning of the Q factor can also change the Purcell enhancement in situ \cite{nakamura_control_2012}. Due to the very small mode volumes, photonic crystal cavities of lower Q factor can reach the same Purcell enhancement as whispering gallery resonators and micropost cavities. This allows faster speeds and higher photon count rates, since ultimately the speed of the device will begin to be limited by the decay rate of the cavity as the spontaneous emission rate of the emitter increases.  In practice, these speeds are difficult to achieve since most single photon sources are pumped via above band optical excitation, whose speed is limited by the modulation rate of the laser.  This is generally fixed at the Ti:Sapphire repetition rate.  Electrical pumping of QDs photonic crystal cavities is challenging but has been implemented \cite{ellis_ultralow-threshold_2011}, and a scheme to optically pump InAs/GaAs QDs with a telecommunications wavelength laser using the intrinsic nonlinearity of the GaAs system to upconvert the light above the band gap has also been implemented.  In this case, the laser can be modulated at GHz speeds using commercially available modulators \cite{rivoire_fast_2011}.  This will be discussed in more detail at the end of section \ref{section:freqconversion}.  Strong coupling has also been obtained for QD-photonic crystal cavity systems \cite{yoshie_vacuum_2004,englund_controlling_2007,hennessy_quantum_2007}.  An SEM and simulated electric field (absolute value) for an L3 photonic crystal cavity \cite{akahane_high-q_2003} are shown in Fig. \ref{fig:microcavities} (c), reproduced from \cite{rivoire_second_2009}.

\subsection{Plasmonic cavities}
\label{section:plasmonic}
Another method (other than TIR and DBR) of confining light to small volumes is using metallic resonators, which exploit plasmonic effects.  Surface plasma oscillations are coherent oscillations of free electrons on a metallic surface.  These charge oscillations are followed by an electromagnetic wave called the surface plasmon
(SP).  Plasmonic devices confine light on thin metallic films or metal particles, into smaller volumes than their dielectric counterparts.  Unfortunately, Q factors are also significantly lower due to the high absorption of metals. Recently, plasmonic \cite{akimov_generation_2007, russell_large_2012, de_leon_tailoring_2012} and hybrid plasmonic-dielectric structures \cite{chen_metallodielectric_2012} have been investigated as another means of enhancing the spontaneous emission rate of emitters.  With metallic nanowires, single colloidal QDs were enhanced over the entire 30 nm inhomogeneously broadened range of emission by a factor of $\sim$2.5 \cite{akimov_generation_2007}. The metallodielectric structure in \cite{chen_metallodielectric_2012} consisted of a defect-free, highly crystalline silver nanowires surrounded by patterned dielectric distributed Bragg reflectors. Experimental Purcell enhancement of up to 75 was demonstrated \cite{chen_metallodielectric_2012} with of an ensemble of colloidal quantum dots (a single QD was not isolated).  In the future this could be extended to single emitters.  Plasmonic enhancement between a silver nanowire and a silver metallic surface \cite{russell_large_2012} showed enhanced spontaneous emission from dye molecules, with a factor of 1000 increase in spontaneous emission rate.  These experiments are made more difficult by the losses in the metals, which lead to high rates of non-radiative recombination.  These losses can also increase the total decay rate, and so care must be taken to determine which part of the decrease in lifetime is due to radiative spontaneous emission rate enhancement and which part is due to an increase in non-radiative emission.

\subsection{Nanowire QDs}
\label{section:nanowires}
Another method for enhancing QD emission and increasing collection efficiency is to use QDs grown in semiconductor nanowire waveguides.  Nanowires are grown on a semiconductor substrate, usually catalyzed with metal nanoclusters.  The nanowires can be subsequently removed from the substrate and placed on another preferrable substrate, or probed while on the substrate.  The QD (or several QDs) is grown within the nanowire. QDs have been demonstrated in nanowires in materials systems such as GaAsP/GaP \cite{borgstrom_optically_2005}, InAsP/InP \cite{reimer_bright_2012}, AlN/GaN \cite{renard_exciton_2008}, CdSe/ZnSe \cite{bounouar_ultrafast_2012} and GaAs/InAs \cite{claudon_highly_2010}.  These nanowires can provide broadband enhancement compared to high Q cavities.  Outcoupling can also be engineered to be very high efficiency.  For example, Claudon et al. demonstrate a collection efficiency of 0.72 over 70 nm bandwidth at 950 nm for InAs/GaAs QDs using a tapered tip and metallic bottom mirror \cite{claudon_highly_2010}, with theoretical spontaneous emission rate 
$\beta$ of up to 0.95.  Reimer et al. \cite{reimer_bright_2012} demonstrated a similar geometry with growth of a single InAsP quantum dot
exactly on the axis of an InP nanowire waveguide.  Further work to develop QD nanowire technology has included the demonstration of high temperature, high-speed \cite{bounouar_ultrafast_2012} and site controlled \cite{tatebayashi_site-controlled_2012} QDs in nanowires.

 \subsection{Efficiency}
 For a single photon source coupled to a microcavity, the quantum efficiency $\eta$ (assuming that all cavity mode photons can be collected) of the single photon source can be viewed as the product of the coupling efficiency of the emitter to the cavity mode $\beta$ and the extraction efficiency of the single photon into a single-mode traveling wavepacket.
For a micropillar cavity in the weak coupling regime $\eta$ is given by
\begin{equation}
\eta = \beta\frac{Q}{Q_{int}}
\end{equation}
where $1/Q_{int}$ is the intrinsic loss due to the finite DBR reflectivity (i.e. $Q_{int}$ is the $Q$ factor of the infinitely wide post) and 1/Q is the total cavity loss. This is a theoretical efficiency, but as mentioned above, experimentally, not all of the light coupled to the cavity will be collected.  A review on photon extraction strategy is given in \cite{barnes_solid-state_2002}.

The coupling of an emitter to the cavity mode is in part determined by the position of the emitter relative to the field maximum (see section \ref{section:strongcoupling}).  Since both spontaneous emission rate and spontaneous emission coupling factor depend on the coupling of the emitter to the cavity mode, the spatial positioning of the emitter with respect to the cavity is very important. Site controlled QDs offer one means for achieving optimal positioning of the emitter by growing QDs at known positions with respect to alignment marks, and subsequently fabricating cavities with reference to these marks \cite{gallo_integration_2008}.  AFM measurements can also reveal the locations of QDs, and a method for positioning relative to measured positions with an accuracy of 10 nm was developed \cite{badolato_deterministic_2005,hennessy_quantum_2007}.  A method of in situ photoluminescence measurement and lithography with an accuracy of $<$ 50 nm was also developed \cite{dousse_controlled_2008}.

\section{Current/future research directions}
\label{section:currentfutureresearch}
    \subsection{Geometries for broadband collection enhancement}

One challenging aspect of using a high-Q microcavity is the necessity for spectral overlap between a narrow cavity mode and the QD emission line. One solution to this problem is to use local tuning techniques to tune the QD in to resonance with the cavity; such methods include temperature \cite{faraon_local_2007}, electrical tuning \cite{faraon_fast_2010}, Zeeman tuning \cite{haft_magneto-optical_2002} and strain tuning \cite{seidl_effect_2006},  or by tuning the cavity resonance, for example via free carrier injection using a strong picosecond pump laser \cite{fushman_ultrafast_2007}, photo-refractive \cite{faraon_local_2008} or liquid nitrogen condensation tuning \cite{mosor_scanning_2005}. Alternatively, spectrally broadband approaches avoid precise tuning.  Photonic crystal waveguides are one way that broadband enhancement can be achieved, in addition to easy on-chip transport and efficient out-coupling.  A photonic crystal waveguide is a line defect in a two dimensional photonic crystal structure.  An emitter coupled to a propagating mode in photonic crystal waveguides can also be subject to broadband Purcell enhancement due to slow-light effects at the edges of the Brillouin zone \cite{hughes_enhanced_2004}. The enhanced emission in the waveguide is shown to scale proportionally with the photon group index.  Predictions of $\beta$ factors of greater than 0.85-0.95 over 10 THz ($\sim$ 40 nm) spectral range  have been predicted for photonic crystal waveguides \cite{rao_single_2007, lecamp_very_2007, yao_controlled_2009}.  Other proposals also consider shorter, finite length photonic crystal waveguides \cite{rao_single_2007}.  These proposals have been confirmed experimentally, with decay enhancements of up to 27 times and $\beta$ factors of up to 0.89 demonstrated \cite{lund-hansen_experimental_2008} over 20 nm.  Similar experiments with different device geometries also saw very high enhancements, $\beta$ factors and bandwidths \cite{patterson_broadband_2009, dewhurst_slow-light-enhanced_2010, thyrrestrup_extraction_2010, schwagmann_-chip_2011}.  The extraction efficiency in the plane of the photonic crystal can be extremely high for these waveguides.  Experimental schemes taking advantage of both the slow light and Fabry-Perot resonance enhancements in short waveguides have also been demonstrated \cite{ba_hoang_enhanced_2012,laucht_waveguide-coupled_2012}.  Other recent experiments used a fiber taper \cite{davanco_efficient_2011} coupled to a waveguide in order to obtain high extraction efficiencies from the waveguide, or a circular dielectric grating pattern around a QD \cite{davanco_circular_2011} for high photon collection efficiencies and lifetime enhancements.  The current work on plasmonic and nanowire structures (discussed in sections \ref{section:plasmonic} and \ref{section:nanowires}) for high efficiency single photon sources also allow more broadband collection. Plasmonic structures can be engineered to have high enhancement over the entire inhomogenous distribution of QDs \cite{akimov_generation_2007}, while nanowires can similarly enhance over a broad spectral range while achieving excellent outcoupling
efficiency \cite{claudon_highly_2010}.

\subsection{Electrically pumped devices-single photon LEDs}
\label{section:electrical}
Electrical injection of a QD can be performed by growing the dot within a p-i-n junction.  Applying a short electrical pulse allows electrons and holes to cross the tops of the barriers and into the QD.  The first example of such a device using a QD grown in a p-n junction \cite{kim_single-photon_1999} used Coulomb blockade for resonant electrical injection; however, very low efficiencies were obtained. Later demonstrations used a micron diameter emission aperture in a high-aluminum content layer to isolate emission from a single QD \cite{Bennett.APL.2005, ellis_oxide-apertured_2007}.  The high-aluminum layer provided both lateral current confinement and helped confine the photonic mode.  An improved device design using a photonic cavity allowed single photon electroluminescence to be demonstrated at repetition rates up to 0.5 GHz \cite{ellis_cavity-enhanced_2008}. Bennet et al. also reported electrical control of the electron and hole populations in a quantum dot.  They showed that with appropriate voltage biasing it is possible to reduce the uncertainty in the time at which photons are emitted from a single quantum dot by an arbitrary factor. By altering the bias across the device, a  time jitter to one-fifth of the radiative lifetime was obtained, and single-photon emission at a repetition rate of 1.07 GHz was observed.  Electrically driven single photon sources have also been demonstrated at different wavelengths and in different QD material systems.  Ward et al. demonstrated an electrically driven 1.3 $\mu$m single-photon source at low temperature \cite{ward_electrically_2007} grown in a planar DBR microcavity, although this source was only operated at 10 MHz.  Triggered single-photon emission from InP/Ga$_{0.51}$In$_{0.49}$P has also been demonstrated in the red spectral range at speeds of up to 200 MHZ \cite{reischle_triggered_2010}.  An electrically-pumped single photon source operating at 1.55 $\mu$m has also been demonstrated \cite{miyazawa_exciton_2008} with InAs/InP QDs.  Recently demonstrated techniques for efficient electrical injection into photonic crystal cavities \cite{ellis_ultralow-threshold_2011} could also be employed in single QD based single photon sources.

    \subsection{Frequency conversion interfaces}
    \label{section:freqconversion}
\begin{figure}
\includegraphics[width = 15cm]{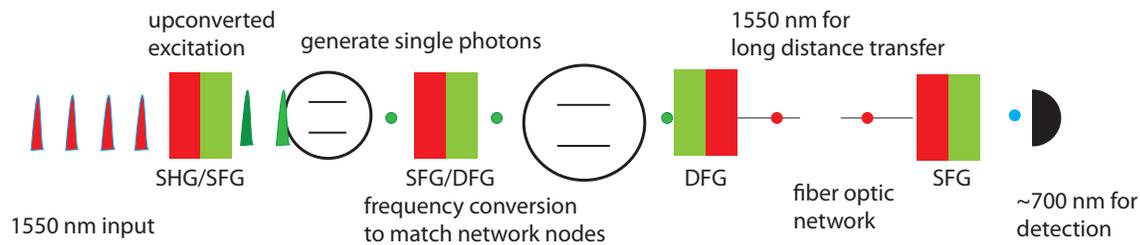}
\caption{A diagram of frequency conversion being used to improve a quantum network.}
\label{fig:freqconversion}
\end{figure}

    QD single photon sources can be fast, stable, efficient and provide good indistinguishability, under either electrical or optical excitation. QD single photon emission has been pushed to higher and higher temperatures in various systems, with room temperature QD emission recently observed.  QDs with single photon emission at telecommunications wavelengths and at visible wavelengths for high detection efficiency have been demonstrated. However, these desirable properties are not all available in the same QD material system.  For interfacing different single photon emitters of different wavelengths, and for interfacing these with the maximum detection/transmission efficiency windows, it is desirable to be able to arbitrarily convert the frequencies of emitted photons.  Theoretically, \cite{kumar_quantum_1990} the quantum nature of photons is preserved during $\chi^{(2)}$ nonlinear frequency conversion processes.  For example, if we take the process of sum-frequency generation in which two light waves of angular frequencies $\omega_1$ and $\omega_2$ are mixed in a nonlinear crystal with a second order nonlinearity to generate the sum frequency $\omega_3 = \omega_1+\omega_2$, given a quantum state such as a single photon input at $\omega_2$ a single photon will also be output at $\omega_3$.  This is the case even when $\omega_2$ is a strong coherent pump, which experimentally will be the usual case in order to obtain the efficiencies needed for the conversion of single photons.  A cartoon of how frequency conversion can be used to improve a quantum network is shown in Fig. \ref{fig:freqconversion}.  A telecommunications wavelength pump is upconverted to pump a QD to emit a single photons.  Two (or more) QDs at different wavelengths can interact as nodes in a quantum network at the single photon level using sum/difference frequency to match their transitions.  The emitted single photons can be converted to telecommunications wavelengths via difference frequency generation for transmission over the fiber optic network.  Finally, sum frequency generation can be used to up-convert the photons for efficient detection on Si APDs.

          The experimental challenge of implementing this scheme is that it requires high efficiency conversion at single photon levels, good in and out-coupling efficiencies and very low noise levels.  Quantum frequency conversion was first observed experimentally in 1992, when Huang et al. demonstrated that when one beam a high flux entangled photon twin beam pump at 1064 nm was upconverted to 532 nm by a strong pump it showed quantum noise reduction indicative of the quantum nature after the upconversion \cite{huang_observation_1992}.  After this, it was several years before the idea was revisited as a means for achieving higher photon detecting efficiency for telecommunications wavelength single photons.

          Quasi-phase matching, a process in which a grating in the nonlinear medium is used to compensate for wave vector mismatch between the input and output wavelengths, has been the most important technology for realizing the near-unity efficiencies required for frequency conversion of single photons. Periodically poled lithium niobate (PPLN), and in particular PPLN waveguides, have been the workhorse for the majority of recent experimental demonstrations.  Unity conversion efficiency can be achieved in these systems.  In practice, the total efficiency is limited by losses in the waveguide and extraction efficiency.  In 2004, two groups \cite{roussev_periodically_2004}, \cite{Albota_efficient_2004} upconverted a weak laser signal at 1340 nm to 720 nm.  An overall conversion efficiency of $>$80\% was demonstrated in a PPLN waveguide and bulk PPLN, using a strong pump laser at 1550 nm. The conversion and detection of a weak laser signal with an overall efficiency of 37 \% was later demonstrated by the Stanford group \cite{langrock_highly_2005}, with greater than 99.9 \% internal efficiency.

          Tanzilli et al. \cite{tanzilli_photonic_2005} demonstrated sum frequency conversion on one of an entangled pair while maintaining the entanglement.  Several down-conversion schemes of a weak laser pump at around 710 nm to telecommunications wavelengths were then shown \cite{takesue_single-photon_2010}, \cite{ding_frequency_2010}, \cite{zaske_efficient_2011}, \cite{ikuta_wide-band_2011} and some major sources of noise affecting these processes were identified \cite{pelc_influence_2010}.  Transduction of a telecommunications band single photon from a quantum dot single photon source was also shown for the first time in 2010, with a 1.3 micron single photon upconverted to 710 nm \cite{Rakher_quantum_2010}.

        A more challenging prospect is to use the intrinsic nonlinearity of the solid state system in which the quantum dot is embedded for frequency conversion.  GaAs, GaP and InP are all non-centrosymmetric crystals with high $\chi^{(2)}$ non-linearities. Highly efficient frequency conversion in optical micro-cavities has been theoretically and experimentally explored, with theoretical schemes for highly efficient quantum dot single photon conversion proposed \cite{mccutcheon_broadband_2009}. Proposals for multiply resonant optical microcavities also show promise for increasing the on-chip frequency conversion efficiencies to the point where frequency conversion at single photon levels is feasible \cite{rivoire_multiply_2011-1, bi_high-efficiency_2012}, and initial realizations of these designs have been fabricated, although conversion of single photons by these microcavities remains in the future.

\begin{figure}
\includegraphics[width = 15cm]{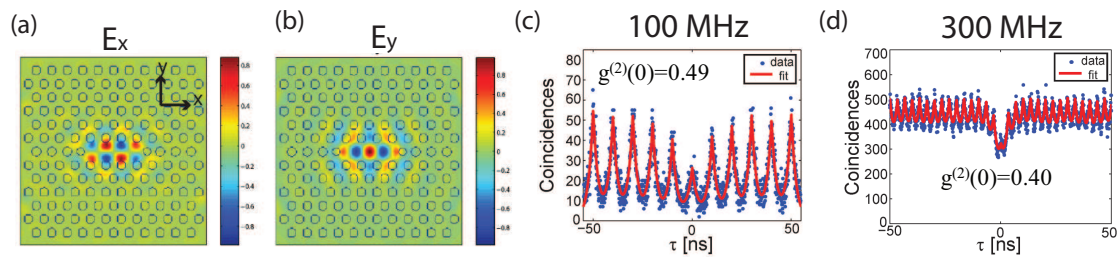}
\caption{A telecommunications wavelength pumped InAs/GaAs QD single photon source.  The telecom laser at around 1500 nm is coupled to the fundamental L3 mode of a photonic crystal cavity, with electric field components in the x and y polarizations shown in (a) and (b).  Intracavity second harmonic is generated at around 750 nm, this excites single InAs QDs.  The laser can be modulated, autocorrelation measurements with the laser modulated at 100 and 300 MHz are shown in (c) and (d).  At 300 MHz the peaks are overlapping significantly, limited by the lifetime of the QD (measured to be 2.4 ns).  Figure parts (a)-(d) reproduced with permission from ref. \cite{rivoire_fast_2011}.}
\label{fig:telecomexc}
\end{figure}

 For interfacing with telecommunications networks, nonlinear excitation forms the other half of the interface. In order to excite InAs/GaAs QDs, above band or resonant excitation is necessary, which means wavelengths shorter than $\sim$ 900 nm.  Interfacing with an optical cavity that can be used to create second harmonic generation which can then excite the QD, all on chip, allows the QD to be directly interfaced with a telecommunications wavelength network, and allows fast modulation via commercially available lithium niobate electro-optic modulators \cite{rivoire_fast_2011}. Using this system, the generation rate of demonstrated optically triggered quantum dot single photon sources was increased above the 80 MHz repetition rate usually used (due to the availability of pulsed Ti:Sapphire lasers at the right wavelength). Electrical excitation in principle allows another way to circumvent this; however, resonant optical excitation improves the indistinguishability of output photons, and many desirable microcavity structures such as photonic crystal cavities have geometries that are challenging to pump electrically.  To determine the maximum speed at which the system could be modulated, an independent experiment directly measuring the dot lifetime was performed, giving a monoexponential decay with time constant 2.4 $\pm$0.1 ns.  This agreed with the maximum rate at which the source could be modulated before coincidences from adjacent peaks began to overlap (300 MHz, see Fig. \ref{fig:telecomexc}).

A doubly resonant cavity with resonances at both telecommunications frequency and the frequency of the dot would increase the spontaneous emission rate of the dot via the Purcell effect, and enable the realization of a significantly faster e.g.,1 GHz triggered single photon source. Such a cavity has been demonstrated in \cite{rivoire_multiply_2011, rivoire_multiply_2011-1}.

    \subsection{Pulse shaping}

    Emitted single photon pulse shapes are generally single sided exponentials decaying with the lifetime of the QD. For InAs/GaAs QDs in bulk, this time is on the order of 1 ns.  For particular applications it is desirable to change the length and shape of this pulse.  Matching pulse length to cavity lifetimes for increasing coupling efficiency, or for storing and retrieving photons from atomic ensembles could benefit from pulse lengthening.  In other cases it may be desirable to shorten pulse lengths, for example for more efficiently sending pulses down fibers. For this application it is also advantageous to make the pulses Gaussian shaped. It has also been shown \cite{rohde_optimal_2005} that Gaussian pulse shapes are optimal for tolerance to mode mismatch. Since one of the major causes of distinguishability between photons is mode mismatch, Gaussian pulses are more desirable for QIP applications where indistinguishable photons are a requirement. Another interesting application of pulse shaping is to observe the response of single emitters to different single photon pulse shapes.

\begin{figure}
\includegraphics[width = 15cm]{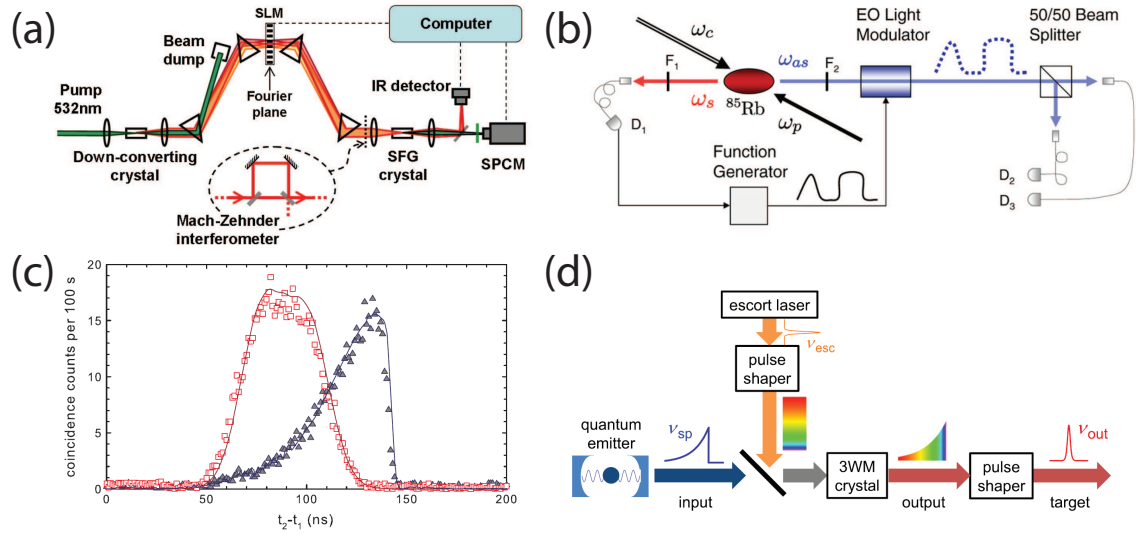}
\caption{(a) Pulse shaping of an entangled photon pair. Reproduced with permission from ref. \cite{peer_temporal_2005}.  (b) Pulse shaping of a single photon via electro-optic light modulator.  (c) Original single sided exponential (triangles) and shaped gaussian (squares) pulse produced via setup in (b). (b) and (c) reproduced with permission from ref. \cite{kolchin_electro-optic_2008}. (d) Proposed scheme for single photon pulse shaping and temporal compression.  Reproduced with permission from ref. \cite{kielpinski_quantum_2011}.}
\label{fig:pulseshaping}
\end{figure}

    In 2005 an experimental demonstration \cite{peer_temporal_2005} of temporal shaping of an entangled photon pair was shown. Entangled photons generated by down-conversion of a continuous pump laser in one crystal were frequency converted via a strong pump to their sum frequency in another crystal.  The pulse shaping was achieved using an arrangement of prisms and lenses to create a spectral Fourier plane.  A computer controlled spatial light modulator applied phase shifts to different spectral components of the entangled photons, before the beam was recombined.  The experimental setup used is reproduced from \cite{peer_temporal_2005} in Fig. \ref{fig:pulseshaping} (a).

    In 2008, temporal shaping of single photons was demonstrated \cite{kolchin_electro-optic_2008, baek_temporal_2008}.  In \cite{kolchin_electro-optic_2008}, a single photon generated via SPDC was shaped using electro-optic modulation.  The electro-optic modulation allowed arbitrary phase and amplitude modulation. One of the entangled pair was used to set the time origin for electro-optic modulation of the wave function of the other photon.  The setup is shown in Fig. \ref{fig:pulseshaping} (b).  Single-photon wave functions with Gaussian shapes, or composed of several pulses were created.  A Gaussian pulse created from a single sided exponential pulse in this experiment is shown in Fig. \ref{fig:pulseshaping} (c).

    Phase shaping of single photon wave packets has also been demonstrated recently \cite{specht_phase_2009}, which for example allows control over interference at a beam splitter.   Very short single photon pulses can also be created via gating with a nonlinear process.  A three wave mixing process will take place when a high power pump is present, so gating a single photon pulse with sum frequency conversion from a very short pump pulse will give very short upconverted pulses  \cite{rakher_simultaneous_2011}. A theoretical scheme \cite{kielpinski_quantum_2011} for for pulse compression via nonlinear mixing with a chirped pump has also been recently proposed. It was shown theoretically that pulse reshaping by this scheme could in principle achieve compression by more than a factor of 100,  with flexible reshaping of the temporal waveform with errors below 1\%.  This scheme is reproduced in Fig. \ref{fig:pulseshaping} (d).

\subsection{Stimulated Raman adiabatic passage (STIRAP)}
\label{section:STIRAP}

\begin{SCfigure}
\includegraphics[width = 10cm]{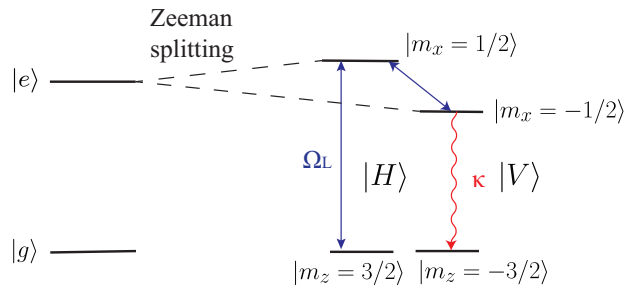}
\caption{QD as a $\Lambda$ system.  A magnetic field applied along the x-axis of a charged QD causes Zeeman splitting (z axis is QD growth axis).  The two transitions are orthogonally linearly polarized.}
\label{fig:stirap}
\end{SCfigure}

The ideal single photon source for quantum information processing would act as both a single photon source and receiver, and could act as a node in a quantum network.  For this ideal source, the single photon emission process must be reversible.  This is not true in the case of incoherent above-band pumping, where the spontaneous emission process is irreversible and cannot be described by a Hamiltonian evolution.  One alternative to incoherent pumping based on a strongly coupled atom-cavity system is stimulated raman adiabatic passage (STIRAP) \cite{parkins_synthesis_1993, law_arbitrary_1996}.  An atom with a  $\Lambda$-system level scheme, shown in Fig. \ref{fig:stirap}, is excited on one transition by a pump laser at the transition frequency $\Omega_L$, stimulating the emission of a single photon into the strongly coupled cavity-emitter transition.

In the case of a QD, a $\Lambda$-system can be implemented by applying a magnetic field to a singly charged QD \cite{imamoglu_quantum_1999, chen_theory_2004}.  When the QD is charged, the lowest energy conduction and valence band states are represented by $|m_x = \pm1/2\rangle$ and $|m_z = \pm3/2\rangle$ respectively, due to the strong z-axis confinement \cite{kiraz_quantum-dot_2004, imamoglu_quantum_1999}. Applying a strong magnetic field along the $x$-axis results in Zeeman splitting of spin states in the conduction band, creating a $\Lambda$-system, shown in Fig. \ref{fig:stirap}.  For an electron Land\'{e} g-factor of 2 and an applied field of 10 T, the splitting is expected to be 1 meV \cite{kiraz_quantum-dot_2004}.  At cryogenic temperatures, this splitting is much larger than the broadenings of the levels.  An x-polarized laser pulse can be applied resonantly between levels $|m_x = 1/2\rangle$ and $|m_z = 3/2\rangle$ or $|m_z = -3/2\rangle$, while levels $|m_x = -1/2\rangle$ and $|m_z = 3/2\rangle$ or $|m_z =-3/2\rangle$ are strongly coupled via a resonant y-polarized cavity mode \cite{kiraz_quantum-dot_2004}.  Limits on the indistinguishability in this scheme include jitter in emission time from spin decoherence of the ground state.

As is necessary for a $\Lambda$ system, direct transitions between the two (almost degenerate) ground states are not allowed.  In order to bring the system from the $|m_z = 3/2\rangle$ state to the $|m_z = -3/2\rangle$ state or vice versa, a laser recycling pulse must be applied \cite{kiraz_quantum-dot_2004}. In the case of a QD, this can simply be a subsequent pulse between the $|m_z = 3/2\rangle$ and $|m_x = 1/2\rangle$ states.  An alternative recycling mechanism can be a Raman $\pi$-pulse generated by two detuned laser pulses satisfying the Raman resonance condition \cite{kiraz_quantum-dot_2004}.

Single photons were generated via STIRAP from Rb \cite{kuhn_deterministic_2002} and Cs \cite{mckeever_deterministic_2004} atoms, with an average of 1.4$\times 10^4$ photons produced from each trapped atom in \cite{mckeever_deterministic_2004}.  Solid state QD $\Lambda$ systems have been implemented, \cite{press_complete_2008, press_ultrafast_2010, ladd_pulsed_2010} however, these experiments focused on spin control of the QD, and did not have a coupled cavity on either transition.  Magnetic tuning of the levels of a charged QD-photonic crystal cavity system \cite{pinotsi_resonant_2011} has been demonstrated, but STIRAP was not demonstrated in this system.

\subsection{Photon blockade}
\label{section:photonblockade}
In the strong coupling regime, the atom and cavity can no longer be thought of as two decoupled systems.  Instead, they must be treated as a single system with an anharmonic ladder of states (Jaynes-Cummings model, see figure \ref{fig:strongcoupling} (b)). Unlike a weakly coupled system, this coupled system must be resonantly excited, and will only accept one photon into the cavity at a time, thereby converting incident coherent light into sub-Poissonian anti-bunched light.  This occurs because after entering the first excited state, it cannot be further excited due to the anharmonicity, i.e. the energy separation between the first and second excited states is different than the separation between the ground state and first excited states.  This phenomenon is analogous to Coulomb blockade, in which charge transport through a device occurs on an electron-by electron basis.  For these strongly coupled systems, the atom cavity system will also now have a much larger cross section than for a single atom.  This effect was first observed for an atom cavity system in 2005 \cite{birnbaum_photon_2005} by measurements of the photon statistics of the transmitted field, and the demonstration of the same effect with two photons via the second excited state was also observed \cite{kubanek_two-photon_2008}. In practice, it can be difficult to observe for solid state systems as the anharmonicity must be greater than the broadening of the energy levels. It was first observed for a quantum dot/photonic crystal cavity system in 2008 \cite{faraon_coherent_2008}. Theoretically the optimum parameters for observing a strong photon blockade have been discussed in \cite{faraon_generation_2010}.  The second manifold of the Jaynes-Cummings ladder was also spectrally resolved recently for a QD cavity system \cite{reinhard_strongly_2012}.  The single photon source may be improved by using a photonic molecule \cite{liew_single_2010,bamba_origin_2011,majumdar_loss-enabled_2012}. Finally, the blockade effect has also been recently observed in microwave superconducting qubit systems \cite{lang_observation_2011}, \cite{hoffman_dispersive_2011}.

\subsection{Detectors}

It is important to note that while the best currently commercially available single photon detectors are based on Si and have maximum detection efficiency in the visible, much research is also going in to improve detectors in the IR.  Superconducting nanowire detectors now show very good efficiencies for single photon detection in telecommunications wavelength, and work on photon number resolving detectors based on this and other technologies is ongoing.  For a review of detectors and their performances, see reference \cite{eisaman_invited_2011}.


\section{Summary}

Single photons are in a distinctly quantum state of light, and entirely different than thermal or coherent light.  Just as the development of the laser and its many applications followed the discovery of stimulated emission, the development of a high efficiency, stable single photon source could enable a whole new class of scientific and engineering endeavors.

Epitaxially grown QDs have been demonstrated in a wide variety of materials systems, and show great promise as a single photon source.  Applications such as quantum information processing (in particular quantum key distribution) and quantum metrology are examples of applications that require single photon sources, albeit with slightly different emphasis on requirements.  For QKD it is important to be able to transmit single photons over long distances at high data rates and with low loss, and to have a g$^{(2)}(0)$ close to zero and a high efficiency in order to maintain security.  On the other hand, linear optical quantum computing requires the ability to interfere single photons and hence very high indistinguishability of the photons is important, in addition to a g$^{(2)}(0)$ close to zero to maintain a low error and high efficiency \cite{knill_scheme_2001}.

For all of these applications, a high data rate is also required in order to make them useful.  The ease of incorporating epitaxially grown QDs into optical microcavities with high spontaneous emission rate enhancement due to the Purcell effect makes them a very attractive system for a fast single photon source. Optical microcavities such as microdisks, microposts and photonic crystal cavities can all be fabricated in the semiconductor substrates in which these QDs are grown. The fastest solid state single photon sources yet demonstrated have been electrically pumped QD sources, demonstrating speeds of up to 1 GHz. That electrical pumping has provided faster speeds than optical pumping is mainly due to the difficulty in optical pumping at speeds above the 80 MHz repetition rate of Ti:Sapphire lasers;  Purcell enhanced lifetimes of $\sim$100 ps have been demonstrated, which in principle (without other technological impediments) should allow speeds of up to 10 GHz.

The indistinguishability of QD single photon sources can also be optimized via the engineering of the spontaneous emission rate.  However, resonant excitation gives the best indistinguishability.  Techniques for resonant excitation of QDs are improving, both using better background suppression techniques and optical cavity effects.  These will lead to higher indistinguishabilities.  Quasiresonant excitation also provides improved indistinguishability (although not as good as resonant), with an easier experimental implementation.

Frequency conversion and pulse shaping of single photons is becoming a more practical technology, in particular using periodically poled lithium niobate waveguides.  The ability to arbitrarily convert emitter wavelengths will allow selection of the optimal wavelength for transmission and detection of single photons.  Additionally, correcting for the discrepancy in the emitter transition energies resulting from inhomogeneous broadening would also be possible, allowing photons from different emitters to be rendered indistinguishable.  Implementation of on-chip single photon frequency conversion in optical microcavities is also promising as an integrated and scalable means for performing frequency conversion of single photons emitted from QDs.

Demonstrations of high temperature single photon sources, and single photon sources at telecommunications wavelengths are exciting, because of their impact on the practicality of these devices.  Improvements in detector technology are also as important in the field, allowing new experiments to be done that could not be done before.  Research on single photon sources also includes the improvement of broadband collection efficiency, photon blockade and improving electrically pumped single photon sources.  Ultimately, nearly perfect indistinguishability could be obtained using a combination of resonant excitation techniques such as adiabatic Raman passage (STIRAP) in the strong coupling regime and by employing a three-level $\Lambda$-system. However, the implementation of a three-level $\Lambda$-system is challenging and requires QD charging and the application of a strong magnetic field, and STIRAP has yet to be demonstrated in a solid state system.

The performance of single photon sources has come a long way since the first demonstration, and so has the performance of QD single photon sources since the first demonstration over 10 years ago.  The first commercially available turn-key single photon source, based on the NV center has recently become available, and it seems likely that fast QD single photon sources will soon follow suit. The ultimate single photon source would be an on-demand, fast, indistinguishable, low error and high efficiency source of single photons, that could be operated cheaply and at high temperature.  It would be narrow linewidth, but tunable over a broad wavelength range.  While many of these qualities have been demonstrated individually in QD single photon sources, engineering a source with all of these qualities has proven a challenge.  For QKD, such a source could have immediate applicability provided the added source security was worth the extra cost.  For formation of
large entangled states via single photon interference, a high efficiency single photon source with these qualities would also lead to an immediate improvement.  In terms of many of the other applications however, the single photon source is just one piece of the puzzle, and more pieces will be needed before these applications can be fully realized.

\section{Acknowledgements}
Financial support was provided by the National Science Foundation (NSF Grant ECCS-
10 25811), Stanford Graduate Fellowships and the NSF GRFP. J.V. would also like to acknowledge funding from the Humboldt Foundation.  S.B. would also like to acknowledge helpful discussions with Michal Bajcsy, Arka Majumdar, Jason Pelc and Tomas Sarmiento.

\section{References}
\bibliographystyle{aip}
\bibliography{QDSPS,Zotero_accents_fixed1,Zotero_accents_fixed2}
\end{document}